\documentclass[12pt]{article}
\usepackage{amssymb}
\usepackage{amssymb}
\usepackage{amssymb}
\usepackage{amssymb}
\usepackage{youngtab}
\usepackage{amsmath,amssymb,latexsym,cite}
\usepackage[]{hyperref}
\input xy
\xyoption{all}

\newcommand{\Sym}{{\rm Sym}}

\newcommand{\mc}{\mathcal}

\begin{document}

\vspace*{0.5in}

\begin{center}

{\large\bf Quantum sheaf cohomology on Grassmannians}

\vspace{0.2in}

Jirui Guo$^1$, Zhentao Lu$^2$, Eric Sharpe$^1$

\vspace*{0.2in}

\begin{tabular}{cc}
{ \begin{tabular}{l}
$^1$ Physics Department \\
Robeson Hall (0435) \\
Virginia Tech \\
Blacksburg, VA  24061
\end{tabular} }
& { \begin{tabular}{l}
$^2$ Mathematical Institute\\
University of Oxford\\
Andrew Wiles Building\\
Radcliffe Observatory Quarter\\
Woodstock Road, Oxford, OX2 6GG
\end{tabular} }
\end{tabular}

{\tt jrkwok@vt.edu}, {\tt zhentao@sas.upenn.edu}, {\tt
ersharpe@vt.edu}

$\,$

\end{center}

In this paper we study the quantum sheaf cohomology of Grassmannians
with deformations of the tangent bundle.  Quantum sheaf cohomology
is a (0,2) deformation of the ordinary quantum cohomology ring,
realized as the OPE ring in A/2-twisted theories.  Quantum sheaf
cohomology has previously been computed for abelian gauged linear
sigma models (GLSMs); here, we study (0,2) deformations of
nonabelian GLSMs, for which previous methods have been intractable.
Combined with the classical result, the quantum ring structure is
derived from the one-loop effective potential. We also utilize
recent advances in supersymmetric localization to compute A/2
correlation functions and check the general result in examples.  In
this paper we focus on physics derivations and examples; in a
companion paper, we will provide a mathematically rigorous
derivation of the classical sheaf cohomology ring.

\begin{flushleft}
December 2015
\end{flushleft}

\newpage

\tableofcontents

\newpage

\section{Introduction}

Computing nonperturbative corrections to charged matter couplings in
heterotic string compactifications is one of the outstanding
problems in string compactifications. On the (2,2) locus, when the
gauge connection is determined by the spin connection, charged
matter couplings such as the ${\bf \overline{27}}^3$ and ${\bf
27}^3$ in compactifications to four dimensions are computed by the A
and B model topological field theories, and their values are by now
well-understood via mirror symmetry. Off the (2,2) locus, much less
is known.

In principle, charged matter couplings off the (2,2) locus can be
computed by the A/2 and B/2 pseudo-topological field theories, and
work has been done in that direction, starting with \cite{4}
(motivated by the mirror symmetry analysis of \cite{abs}). These
twists of a (0,2) nonlinear sigma model on a space $X$ with bundle
${\cal E}$ exist when
\begin{displaymath}
\det {\cal E} \: \cong \: K_X^{\pm 1}, \: \: \: {\rm ch}_2({\cal E})
\: = \: {\rm ch}_2(TX).
\end{displaymath}
OPE's in these pseudo-topological theories define `quantum sheaf
cohomology' rings, generalizing ordinary quantum cohomology rings.
For example, in A/2 theories, the chiral states are of the form
\begin{displaymath}
\oplus H^{\bullet}\left(X, \wedge^{\bullet} {\cal E}^* \right),
\end{displaymath}
and quantum sheaf cohomology encodes nonperturbative quantum
corrections to the product structure on the sheaf cohomology above.
This work continued in {\it e.g.}
\cite{gk,mcom,s2,s3,ade,tan1,tan2,ms,m1,kmmp,gs2,mp1,pmp,mcorev,g1,mm3},
and culminated in a description of quantum sheaf cohomology rings on
toric varieties with gauge bundles given by deformations of the
tangent bundle, as described physically in GLSMs in \cite{5,6} and
mathematically in \cite{dgks1,dgks2}. (See also
\cite{mss,gs,dlm,sm14,ang1,lu1,cgjs} for more recent discussions,
and \cite{bbcl} for a recent discussion of perturbative
contributions to Yukawa couplings.)

Although those results are an important step, computing
nonperturbative corrections and quantum sheaf cohomology for compact
Calabi-Yau's with bundles that are not deformations of tangent
bundles remains an open question.

As a stepping-stone towards that goal, we have been considering
quantum sheaf cohomology on Grassmannians.  These have technical
complications beyond those of toric varieties, yet also have enough
symmetries to make one hope that a tractable solution exists. In
terms of GLSMs, this involves understanding nonabelian cases,
whereas all previous work in quantum sheaf cohomology in (0,2)
models has been in abelian GLSMs. In terms of the underlying
mathematics, this becomes a story about nontrivial sheaves on Quot
schemes, a technical step beyond toric cases, for which the
pertinent moduli spaces are again toric varieties and induced
sheaves are locally-free.

In this paper, we will present the results of that program, namely
quantum sheaf cohomology rings on Grassmannians with deformations of
the tangent bundle. Specifically, we will use one-loop effective
action and supersymmetric localization \cite{cgjs,7,7a,8,8a}
arguments to derive and discuss the form of the quantum sheaf
cohomology ring.
For generic deformations off the (2,2) locus, we will argue that
the quantum sheaf cohomology ring can be expressed as
\begin{displaymath}
\begin{array}{l} 
{\mathbb C}\left[\sigma_{(1)}, \sigma_{(2)}, \cdots  \right] /
\left\langle D_{k+1}, D_{k+2}, \cdots, R_{(n-k+1)}, \cdots,
R_{(n-1)},
\right. \\
\hspace*{1.5in} \left. R_{(n)}+q, R_{(n+1)} + q \sigma_{(1)},
R_{(n+2)} + q \sigma_{(2)}, \cdots
 \right\rangle,
\end{array} 
\end{displaymath}
where
\begin{eqnarray*}
D_m & = & \det\left(\sigma_{(1+j-i)} \right)_{1 \leq i,j \leq m}, \\
R_{(r)} & = & \sum_{i=0}^{ {\rm min}(r,n) } I_i \sigma_{(r-i)}
\sigma_{(1)}^i,
\end{eqnarray*}
for $I_i$ the coefficients of the characteristic polynomial of a matrix
$B$ defining the deformation.
We discuss how the ring above encodes the ordinary quantum
cohomology ring (and the classical cohomology ring) of the
Grassmannian as special cases.  
We also discuss two sets of non-generic loci where the expression above
fails to hold.  Along one, the 
`discriminant locus', both the deformed bundle and the corresponding
physical theory degenerate.  We give explicit expressions for this
locus.  In addition, there is a second
locus of interest,
where the additive part of the cohomology ring jumps.  We derive
an expression for this second `jumping' locus.

In a companion paper \cite{3}, we
will give a mathematical proof of the classical sheaf cohomology
ring corresponding to the $q \rightarrow 0$ limit of our quantum
sheaf cohomology ring. A purely mathematical proof of the form of
the quantum sheaf cohomology ring on Grassmannians with deformations
of the tangent bundle, an exercise in sheaf theory on Quot schemes,
is left for the future.

The paper is organized as follows. In section~\ref{sect:basicphys},
we discuss general issues regarding (0,2)-deformations on the
Grassmannian, its one-loop effective potential on the Coulomb branch
and supersymmetric localization. In section~\ref{sect:qsc-ring}, we
give a representation for the quantum sheaf cohomology by making use
of our result in \cite{3} and the one-loop effective potential. We
also discuss its applicability:  our formula will be valid for
generic deformations, but will break down on certain codimension one
subvarieties, which we discuss in detail.  We also check that our
quantum sheaf cohomology ring correctly specializes to both the
classical and quantum (ordinary) cohomology rings, as expected on
the (2,2) locus where the bundle is just the tangent bundle. In
section~\ref{examples}, we check the given quantum sheaf cohomology
ring in examples.  We compute correlation functions using
supersymmetric localization, yielding analogues of
Jeffrey-Kirwan-Grothendieck residues on the Coulomb branch, as in
\cite{cgjs}.  We also explicitly discuss the codimension-one
subvarieties along which our description of the quantum sheaf
cohomology ring breaks down. In appendix~\ref{app:math-rep}, we give
a mathematical description of the classical sheaf cohomology ring,
outlining the approach and results of \cite{3}.  In
appendix~\ref{app:hom-alg}, we outline how the product structures
can be understood as an exercise in homological algebra, leading to
a speculation that `quantum sheaf cohomology' might be understood as
`quantum homological algebra.'  Finally, in
appendix~\ref{app:trivial-defs}, we outline the arguments that will
appear in \cite{3} in the special case of a deformation defined by
$B \propto I$, for which the resulting bundle is isomorphic to the
tangent bundle, and the deformation trivial.

In passing, the reader should note that when we speak of 
``(0,2) chiral rings'' or OPE rings, we are referring to a 
finite-dimensional truncation of the infinite-dimensional chiral ring
of a (0,2) theory, a truncation which reduces on the (2,2) locus to the
ordinary (2,2) chiral ring.  It was argued in \cite{ade} that the OPE
algebra of this truncation closes into itself, so it is consistent to
refer to this as an OPE ring.  In any event, this is the ring in the
twisted theory that
physically defines quantum sheaf cohomology.

\section{Nonabelian A/2 models}  \label{sect:basicphys}

\subsection{(0,2) deformation}

The gauged linear sigma model can be used to implement various
geometric settings. On the (2,2) locus, the Grassmannian $G(k,n)$ is
described by a two-dimensional supersymmetric $U(k)$ gauge theory
with $n$ chirals in the fundamental representation. Let's denote
these chiral fields by $\Phi^i_\alpha, \alpha=1,\cdots,k,
i=1,\cdots,n$. The (2,2) vector supermultiplet decomposes into a
(0,2) vector multiplet $V$ and a chiral multiplet $\Sigma$. The
bosonic component of $\Sigma$ is an adjoint valued scalar $\sigma$.
The chiral supermultiplet, $\Phi^i_a$, decomposes into a (0,2)
chiral multiplet $\Phi^i_\alpha = (\phi^i_\alpha, \psi^i_{+
\alpha})$ and a (0,2) Fermi multiplet $\Lambda^i_\alpha = (\psi^i_{-
\alpha}, F^i_\alpha)$, obeying
\begin{equation}\label{1}
\overline{D}_+ \Lambda^i_\alpha \: = \: \sigma^\beta_\alpha
\Phi^i_\beta .
\end{equation}
In a (0,2) theory, the covariant derivative of the Fermi superfield
can be any function annihilated by the covariant derivative, i.e.,
\eqref{1} is generalized to
\begin{equation}\label{2}
\overline{D}_+ \Lambda^i_\alpha \: = \: E^i_\alpha,
\end{equation}
where $E$ is a holomorphic function of the chiral superfields
satisfying
\[
\overline{D}_+ E = 0.
\]
In particular, we can deform off the (2,2) locus by taking
\begin{displaymath}
\overline{D}_+ \Lambda^i_\alpha \: = \:A^i_j \sigma^\beta_\alpha
\Phi^j_\beta \: + \: B^i_j ({\rm Tr} \, \sigma) \Phi^j_\alpha,
\end{displaymath}
where $A$ and $B$ are $n$ by $n$ matrices.
For simplicity, in this paper we will assume $A$ is invertible,
which will guarantee our models can be deformed to the (2,2) locus.
In principle, one could also imagine nonlinear deformations, functions of
say
\begin{displaymath}
\epsilon^{\alpha_1 \cdots \alpha_k} \Phi^{i_1}_{\alpha_1} \cdots
\Phi^{i_k}_{\alpha_k},
\end{displaymath}
but as conjectured in \cite{mcom} and later demonstrated in
\cite{dgks1,dgks2,dlm,cgjs}, 
the A/2 model correlation functions and quantum sheaf cohomology
ring relations are independent of nonlinear deformations, so we only 
consider linear deformations.
\par
The left moving fermion is now a section of the vector bundle
$\phi^* \mathcal{E}$, where $\mathcal{E}$ is a vector bundle on
$G(k,n)$ defined by the short exact sequence
\begin{equation}
0 \to \mathcal{S} \otimes \mathcal{S}^* \stackrel{g}{\to}
\mathcal{V} \otimes \mathcal{S}^* \to \mathcal{E} \to 0,
\label{deformed1}
\end{equation}
where $g$ can be represented as
$$\omega^{\beta}_{\alpha} \mapsto
A^i_j \omega^{\beta}_{\alpha} x^j_{\beta} + \omega^{\beta}_{\beta}
B^i_j x^j_{\alpha}.$$
 The dual of
\eqref{deformed1} is
\begin{equation}
0 \to \mathcal{E}^* \stackrel{i}{\to} \mathcal{V}^* \otimes
\mathcal{S} \stackrel{f}{\to} \mathcal{S}^* \otimes \mathcal{S} \to
0, \label{deformed2}
\end{equation}
where $f$ can be represented as
$$t^{\alpha}_i \mapsto t^{\alpha}_i
f^i_{\beta} = t^{\alpha}_i A^i_j x^j_{\beta} +
\delta^{\alpha}_{\beta} t^{\gamma}_i B^i_j x^j_{\gamma}.$$ Our goal
is to study the quantum sheaf cohomology ring
$$\underset{r\geqslant 0}{\oplus} H^r(G(k,n), \wedge^r \mathcal{E}^*).$$

The number of bundle moduli is equal to $h^1(X, {\rm End}~TX)$. In
the case at hand, $X=G(k,n)$, and $TX=\mathcal{S}^* \otimes
\mathcal{Q}$, where $\mathcal{S}$ is the universal vector bundle and
$\mathcal{Q}$ is the universal quotient bundle. Applying the
Borel-Weil-Bott theorem, one can compute
\begin{displaymath}
 h^1(G(k,n), {\rm End}~TG(k,n)) \: = \:
\left\{ \begin{array}{cl}
n^2-1 & 1 < k < n-1, \\
0 & {\rm else}.
\end{array} \right.
\end{displaymath}
In other words, projective spaces have no tangent bundle moduli, but
other Grassmannians do. Let us see how this number emerges from our
description of the deformation.\par

Our description above encodes moduli in the two $n \times n$
matrices $A$, $B$. The invertible matrix~$A$~can be transformed into
the identity matrix using a $GL(n)$ field redefinition,
so that in
effect only one matrix ($B$, or rather $B A^{-1}$) encodes the
moduli.  However, the overall trace in $B$ is trivial, and does not
define any bundle deformations, which we can see as follows.
Without loss of generality, take $A$ to be the identity. Denote by
$i$ the imbedding of $\mathcal{S}$ in $\mathcal{V}$. Given a local
section of $\mathcal{V}^* \otimes \mathcal{S} = \mathcal{H} om
\mathcal{(V,S)}$, denoted by $t$,
 $f(t)$ can be written as $ti+{\rm Tr} (tBi) I_{k\times k}$,
where $I_{k\times k}$ is the $k \times k$ identity matrix. If $t$ is
in the kernel of $f$ and $B=\varepsilon I_{n \times n}$, then
\begin{equation}\label{ti}
ti + \varepsilon~ {\rm Tr}(ti)~ I_{k\times k} = 0.
\end{equation}
Taking the trace, we get
\[
(1+\varepsilon k)~ {\rm Tr}(ti)=0.
\]
For generic~$\varepsilon$, this implies~${\rm Tr}(ti)=0$, but
then~$ti=0$~by \eqref{ti}. This means~$t$~is in the kernel of~$f_0$
($f$ with $B=0$). The converse is also true. We conclude that
$\mathcal{E}^* \cong  \Omega$, the holomorphic cotangent bundle,
when $B=\varepsilon I_{n \times n}$. Thus, we see the number of
nontrivial deformations is $n^2-1$, encoded in $B$ (or $B A^{-1}$ if
$A$ is nontrivial), modulo an overall trace.

Not all $n \times n$ matrices define a vector bundle through
equation~\eqref{deformed1}. In fact, in \cite{3}, we show that a
$B$-deformation fails to give rise to a vector bundle on $G(k,n)$ if
and only if there exist $k$ eigenvalues of $B$ (or $B A^{-1}$, if
$A$ is nontrivial) that sum to $-1$. Physically, if this condition is
satisfied, then the GLSM develops a noncompact branch, independent of
the value of the Fayet-Iliopoulos parameter.  In any event,
this criterion gives us the
discriminant locus along which the $A/2$ correlation functions
diverge.

\subsection{One-loop effective potential}

We will derive the quantum sheaf cohomology ring relations from the
one-loop effective potential on the Coulomb branch, which we review
in this section.

For the GLSM corresponding to $G(k,n)$, the gauge group is $U(k)$,
which is generically\footnote{
For our computations, we will be able to essentially
ignore loci with enhanced
gauge symmetry.  For example, in supersymmetric localization computations,
residues vanish along such loci, because of {\it e.g.}
factors of the form
\begin{displaymath}
\prod_{a \neq b} \left( \sigma_a - \sigma_b \right)
\end{displaymath}
in the numerator of the integrand.
As a result, such loci do not contribute to our computations, and will
be ignored in this paper.
}
broken to $U(1)^k$ along the Coulomb branch.  For $\sigma$
the adjoint-valued field in the (2,2) vector multiplet, Take
$\sigma_a, a=1,\cdots,k,$ to be the components of $\sigma$ in the
Cartan subalgebra.  These will act as coordinates along the Coulomb
branch. On this branch, the charge for $\Phi^i_a$ is $\delta^b_a$
under the $b$-th $U(1)$. Notice that all the $\Phi^i_a$'s with the
same $a$ have the same charges under all the $U(1)$'s. For fixed
$a$, we can rewrite \eqref{2} as
\[
\overline{\mathcal{D}}_+ \Lambda^i_a = E^i_{j}(\sigma_{a}) \Phi^j_a,
\]
where the $n \times n$ matrix $E^i_j$ is given by
\begin{displaymath}
E^i_j(\sigma_a) = \sigma_a A^i_j + {\rm Tr}(\sigma) B^i_j
\end{displaymath}
for general $A$, or for $A$ taken to be the identity,
\begin{displaymath}
E^i_j(\sigma_a) = \sigma_a \delta^i_j  + {\rm Tr}(\sigma) B^i_j.
\end{displaymath}
According to \cite{4}, the one-loop effective $J$ function is
\begin{equation}\label{J(0,2)}
\tilde{J}_a = - \ln\left[ -q^{-1} \det( E_a )\right].
\end{equation}
(Here a minus sign is inserted to comply with the convention in
mathematical literature, this corresponds to an overall shift in the
theta angle.) The equations of motion are $\tilde{J}_a = 0$ for each
$a$, or more simply, for each $a$,
\begin{equation}\label{emo}
\det(E(\sigma_a)) = \det( \sigma_a A + {\rm Tr}(\sigma) B ) = -q
\end{equation}
for general matrices $A$.

\subsection{Supersymmetric localization}

We shall check the predictions for the quantum sheaf cohomology ring
by computing correlation functions in examples, using supersymmetric
localization.  Now, it is not known how to apply supersymmetric
localization to an untwisted (0,2) theory, but in this paper we are
concerned with a twist of the (0,2) theory, known as the A/2 model.
In a (0,2) nonlinear sigma model on a space $X$ with bundle ${\cal
E}$, we can understand the A/2 twist as follows. Before the twist,
the right moving fermion $\psi_+$ is a section of $K^{1/2}\otimes
\phi^*TX$, and the left moving fermion $\lambda_-$ is a section of
$\overline{K}^{1/2} \otimes \phi^* \mathcal{E}^*$, where $K$ is the
canonical line bundle of the worldsheet. In an A/2 twisted nonlinear
sigma model \cite{4}, for example, we have
\[
\begin{split}
&\psi_+^i \in \Gamma(\phi^* T^{1,0}X), \\
&\psi_+^{\bar{\imath}} \in \Gamma(K \otimes \phi^* T^{0,1}X),\\
&\lambda_-^a \in \Gamma(\overline{K} \otimes \phi^* \mathcal{E}^*),\\
&\lambda_-^{\bar{a}} \in \Gamma(\phi^* \overline{\mathcal{E}}^*),
\end{split}
\]
with the chiral ring being isomorphic to
$$\underset{r\geqslant 0}{\oplus} H^r(X, \wedge^r \mathcal{E}^*).$$

In the UV GLSM for the Grassmannian $G(k,n)$, the gauge-invariant
chiral ring operators are of the form ${\rm Tr} \, \sigma^k$ for
integers $k$ and $\sigma$ the 
bosonic
field of the chiral multiplet in the adjoint representation. We will express
these in terms of symmetric polynomials in commuting elements forming
a basis along the Coulomb branch, denoted $\sigma_a = 
\sigma_1, \sigma_2, \cdots,\sigma_k$.  A/2 correlation functions of 
symmetric polynomials
in the $\sigma_a$ then
suffice to
determine the quantum sheaf cohomology associated with the chiral
ring. In terms of these commuting elements,
the bosonic potential becomes of the form
\begin{eqnarray*}
\lefteqn{ \sum_{i, a} | A^i_j \sigma_a \phi^j_a \: + \: B^i_j
({\rm Tr} \, \sigma) \phi^j_a |^2
} \\
& = & \sum_{i, a} \overline{\phi}^j_a \phi^k_a \left( A^i_j
\sigma_a \: + \: B^i_j ({\rm Tr} \, \sigma)  \right)^*
\left( A^i_k \sigma_a \: + \: B^i_k ({\rm Tr} \, \sigma)
 \right)
\\
& = & \sum_{i, a} \overline{\phi}^j_a \phi^k_a ( E^{i}_{j, a} )^*
E^{i}_{k, a}
\end{eqnarray*}
where
\begin{displaymath}
E^{i}_{j, a} \: = \: A^i_j \sigma_a \: + \: B^i_j ({\rm Tr} \,
\sigma) 
\end{displaymath}
The Yukawa couplings have the form
\begin{displaymath}
- \overline{\psi}^j_{- a} \psi^i_{+ a} E^{i}_{j, a} \: + \: {\rm
c.c.}.
\end{displaymath}
These couplings -- the bosonic potential and Yukawa couplings --
define what amount to $\sigma$-dependent masses that play a crucial
role in the one-loop partition function in supersymmetric
localization.

Supersymmetric localization in the A/2 model for (0,2) theories
given by deformations of (2,2) theories was recently discussed in
\cite{cgjs}.  From the results there,
\begin{displaymath}
Z^{\rm 1-loop} \: = \:  \prod_{a=1}^k \left( \frac{1}{ \det
\tilde{E}(\sigma_{a}) } \right)
\end{displaymath}
where
\begin{displaymath}
\tilde{E}^i_j(\sigma_{a}) \: = \: A^i_j \sigma_{a} \: + \:
B^i_j \left( \sum_{b} \sigma_{b} \right)
\end{displaymath}
This implies that for any polynomial $f$ in $\sigma_a, a=1,\cdots,
k$, the correlation functions off the (2,2) locus should have the
form
\begin{equation}\label{localization(0,2)}
\begin{split}
&\langle f(\sigma) \rangle =\\  & \frac{1}{k!} \sum_{{\mathfrak
m}_1,\cdots,{\mathfrak m}_k \in {\mathbb Z} } {\rm JKG-Res} \,
\left\{ (-1)^{(n-1) \sum {\mathfrak m}_i} q^{\sum {\mathfrak m}_i}
\left( \prod_{a \neq b} (\sigma_{a} - \sigma_{b} )
\right) \prod_{a=1}^k \left( \frac{1}{\det
\tilde{E}(\sigma_{a})} \right)^{{\mathfrak m}_i+1} f(\sigma)
\right\},
\end{split}
\end{equation}
where `JKG' denotes the Jeffrey-Kirwan-Grothendieck residues defined
in \cite{cgjs}.

In principle, given the $A/2$ correlation functions, the quantum
sheaf cohomology ring is defined in the same way as the ordinary
quantum cohomology. If we take a basis $e_i$ for
$\underset{r\geqslant 0}{\oplus} H^r(G(k,n), \wedge^r
\mathcal{E}^*)$ as a vector space, and a dual basis $\hat{e}_i$ in
the sense that
\[
\langle e_i \hat{e}_j \rangle = \delta_{i j},
\]
the generating relations read
\[
\sigma = \sum_i \langle \sigma e_i \rangle \hat{e}_i
\]
for any $\sigma$. More to the point, the quantum (sheaf) cohomology
ring relations define identities in the correlation functions:  if
in the ring, some quantity $R$ is set to zero, then any correlation
function containing $R$ should vanish. We will use localization to
check the ring structure in examples in section~\ref{examples}.

\section{Ring structure of quantum sheaf cohomology}
\label{sect:qsc-ring}

The quantum sheaf cohomology ring is the OPE ring of an A/2-twisted
theory, just as the ordinary quantum cohomology ring is the OPE ring
of an A-twisted theory -- quantum sheaf cohomology is the (0,2)
generalization of ordinary quantum cohomology.  In this section we
will describe it for Grassmannians with deformations of the tangent
bundle, and give a physics-based derivation.

Also, so far we have given results for general deformation matrices
$A$ and $B$, but as previously observed, the matrix $A$ is
redundant. In the rest of this paper, we will assume without loss of
generality that $A$ is the identity.  The general case can be
reconstructed by replacing $B$ (in results derived for $A=I$) with
$B A^{-1}$.

\subsection{Gauge-invariant operators}
\label{sect:gauge-inv-ops}

The Coulomb branch arguments given in the last section, both
one-loop effective actions and supersymmetric localization, involve
for a $U(k)$ gauge theory a set of $k$ mutually commuting fields
$\sigma_1, \cdots, \sigma_k$ which act as local coordinates on the
Coulomb branch.  However, these individually are not quite invariant
under $U(k)$, as there is still a residual Weyl group action.

The complete group invariants are symmetric polynomials in
$\sigma_1, \cdots, \sigma_k$, and these can be naturally associated
to Young diagrams, via what are known as Schur polynomials (see
\cite{fulton}[chapter 6] or
\cite{jsw}[appendix B] for an introduction).  For example, if $k=2$,
then
\begin{eqnarray*}
\sigma_{\tiny\yng(1)} & = & \sigma_1 + \sigma_2, \\
\sigma_{\tiny\yng(2)} & = & \sigma_1^2 + \sigma_2^2 + \sigma_1 \sigma_2, \\
\sigma_{\tiny\yng(3)} & = & \sigma_1^3 + \sigma_1^2 \sigma_2 +
\sigma_1 \sigma_2^2 + \sigma_2^3, \\
\sigma_{\tiny\yng(1,1)} & = & \sigma_1 \sigma_2, \\
\sigma_{\tiny\yng(2,1)} & = & \sigma_1^2 \sigma_2 + \sigma_1 \sigma_2^2, \\
\sigma_{\tiny\yng(2,2)} & = & \sigma_1^2 \sigma_2^2,
\end{eqnarray*}
and so forth.  Each polynomial is homogeneous, of degree equal to
the number of boxes in the Young diagram.

As we will see in detail in appendix~\ref{app:math-rep} and
\cite{3}, Young diagrams as above correspond mathematically to
elements of sheaf cohomology groups
\begin{displaymath}
H^{\bullet}\left( G(k,n), \wedge^{\bullet} {\cal E}^* \right),
\end{displaymath}
for ${\cal E}$ the pertinent tangent bundle deformation, which arise
in nonlinear-sigma-model-based analyses.  For example, there is a
well-known correspondence between generators of cohomology of the
Grassmannian $G(k,n)$ of fixed degree, and Young diagrams that fit
inside a $k \times (n-k)$ box.  (Young diagrams that extend outside
of that box would correspond mathematically to cohomology classes of
too-high degree, which classically vanish.)

Now, for the purposes of describing the ring, including all the
Young diagrams is redundant, as there are relations between their
products.  For example, in the $k=2$ case above,
\begin{displaymath}
\sigma_{\tiny\yng(1)}^2 \: = \: \sigma_1^2 + 2 \sigma_1 \sigma_2 +
\sigma_2^2 \: = \: \sigma_{\tiny\yng(2)} + \sigma_{\tiny\yng(1,1)},
\end{displaymath}
and so $\sigma_{\tiny\yng(1,1)}$ is determined algebraically by
$\sigma_{\tiny\yng(1)}^2$ and $\sigma_{\tiny\yng(2)}$.  More
generally, the symmetric polynomials corresponding to any Young
diagram that extends past the first row can be expressed
algebraically in terms of Young diagrams that run along the first
row only.  This is known as the Giambelli formula (see {\it e.g.}
\cite{fulton}[section 9.4]), which for a Young diagram $\lambda$,
reads
\begin{displaymath}
\sigma_{\lambda} \: = \: \det \left( \sigma_{(\lambda_i + j-i) }
\right)_{ 1 \leq i, j \leq r}
\end{displaymath}
for $r$ the number of boxes in $\lambda$, $\lambda_i$ the number of
boxes in the $i$th row, and $\sigma_{(n)}$ corresponding to a Young
diagram with one horizontal row of $n$ boxes, {\it e.g.}
\begin{displaymath}
\sigma_{(1)} \: = \: \sigma_{\tiny\yng(1)}, \: \: \: \sigma_{(2)} \:
= \: \sigma_{\tiny\yng(2)}, \: \: \: \sigma_{(3)} \: = \:
\sigma_{\tiny\yng(3)},
\end{displaymath}
and so forth, in conventions in which $\sigma_{(m)} = 0$ for $m <
0$, and is $1$ if $m=0$. For example, the Giambelli formula says
\begin{displaymath}
\sigma_{\tiny\yng(1,1)} \: = \: \det \left[
\begin{array}{cc}
\sigma_{\tiny\yng(1)} & \sigma_{\tiny\yng(2)} \\
1 & \sigma_{\tiny\yng(1)} \end{array} \right] \: = \:
\sigma_{\tiny\yng(1)}^2 \: - \: \sigma_{\tiny\yng(2)},
\end{displaymath}
which we verified explicitly above. For another example,
\begin{displaymath}
\sigma_{\tiny\yng(2,1)} \: = \: \det \left[ \begin{array}{ccc}
\sigma_{\tiny\yng(2)} & \sigma_{\tiny\yng(3)} & \sigma_{\tiny\yng(4)} \\
1 & \sigma_{\tiny\yng(1)} & \sigma_{\tiny\yng(2)} \\
0 & 0 & 1
\end{array} \right] \: = \: \sigma_{\tiny\yng(1)}
\sigma_{\tiny\yng(2)} \: - \: \sigma_{\tiny\yng(3)},
\end{displaymath}
which is easily checked.

Altogether, the classical cohomology ring of the Grassmannian
$G(k,n)$ can be expressed in terms of generators corresponding to
Young diagrams with only a single horizontal row, as
\cite{buch1,buch2,tamvakisrev,bertram1,edver,st}
\begin{equation}  \label{eq:ord-cohom-grassmannian}
{\mathbb C}\left[\sigma_{(1)}, \cdots, \sigma_{(n-k)}\right] /
\left\langle D_{k+1}, \cdots, D_{n} \right\rangle,
\end{equation}
where
\begin{displaymath}
D_m \: = \: \det\left(\sigma_{(1+j-i)} \right)_{1 \leq i,j \leq m},
\end{displaymath}
in conventions in which $\sigma_{(m)} = 0$ if $m<0$ or $m>n-k$, as
each $D_m$ should only be constructed from the available generators.

It should be mentioned that the classical cohomology ring can also
be written in the presentation
\begin{displaymath}
{\mathbb C}\left[ \sigma_{(1)}, \cdots, \sigma_{(k)} \right] /
\left\langle D_{n-k+1}, \cdots, D_n \right\rangle.
\end{displaymath}
These two presentations define equivalent rings.  When describing
the ordinary cohomology ring of the Grassmannian, it is often
convenient to think of the generators as Chern classes:  Chern
classes of the universal quotient bundle in
(\ref{eq:ord-cohom-grassmannian}, and Chern classes of the universal
subbundle above (see {\it e.g.} \cite{edver}).  For the ordinary
cohomology ring, they can be related by transposing Young diagrams,
though that description is not symmetric in the quantum case. In any
event, in this paper we will primarily refer back to
presentation~(\ref{eq:ord-cohom-grassmannian}).

Now, we are interested in computing both quantum and (0,2)
modifications to the classical Grassmannian cohomology ring
structure, so {\it a priori}, it might happen that Young diagrams
extending past the first row are needed. Nevertheless, it turns out
that they are not, the quantum sheaf cohomology ring can be
determined solely by Young diagrams along the first row only.

As a special case, the standard result for the ordinary quantum
cohomology ring of $G(k,n)$ is ({\it e.g.}
\cite{buch1,buch2,tamvakisrev,bertram1,edver,st})
\begin{displaymath}
{\mathbb C}\left[ \sigma_{(1)}, \cdots, \sigma_{(n-k)} \right] /
\left\langle D_{k+1}, \cdots, D_{n-1}, D_n - (-)^{n-k-1} q
\right\rangle,
\end{displaymath}
or, in terms of the other presentation of the classical cohomology
ring,
\begin{displaymath}
{\mathbb C}\left[ \sigma_{(1)}, \cdots, \sigma_{(k)} \right] /
\left\langle D_{n-k+1}, \cdots, D_n - (-)^{k-1} q \right\rangle.
\end{displaymath}

\subsection{Quantum sheaf cohomology ring}
\label{sect:qsc-ring-details}

We will see that the quantum sheaf cohomology ring (the OPE ring of
the A/2 twist) of a (0,2) deformation of the Grassmannian $G(k,n)$
is given {\it generically} by
\begin{equation}   \label{eq:general-qsc-ring}
\begin{array}{l}
{\mathbb C}\left[\sigma_{(1)}, \sigma_{(2)}, \cdots  \right] /
\left\langle D_{k+1}, D_{k+2}, \cdots, R_{(n-k+1)}, \cdots,
R_{(n-1)},
\right. \\
\hspace*{1.5in} \left. R_{(n)}+q, R_{(n+1)} + q \sigma_{(1)},
R_{(n+2)} + q \sigma_{(2)}, \cdots
 \right\rangle,
\end{array}
\end{equation}
specializing for $k=1$ to
\begin{displaymath}
{\mathbb C}\left[ \sigma_{(1)}, \sigma_{(2)}, \cdots  \right] /
\left\langle D_{k+1}, D_{k+2}, \cdots,  R_{(n)} + q , R_{(n+1)} + q
\sigma_{(1)}, \cdots \right\rangle,
\end{displaymath}
and for $k=n-1$ to
\begin{displaymath}
{\mathbb C}\left[ \sigma_{(1)}, \sigma_{(2)}, \cdots  \right] /
\left\langle
 D_{n}, D_{n+1}, \cdots, R_{(2)}, \cdots, R_{(n-1)},
R_{(n)} + q, R_{(n+1)} + q \sigma_{(1)}, \cdots \right\rangle,
\end{displaymath}
where
\begin{eqnarray*}
D_m & = & \det\left(\sigma_{(1+j-i)} \right)_{1 \leq i,j \leq m}, \\
R_{(r)} & = & \sum_{i=0}^{ {\rm min}(r,n) } I_i \sigma_{(r-i)}
\sigma_{(1)}^i,
\end{eqnarray*}
for $I_i$ the coefficients of the
characteristic polynomial of $B$, given by
\begin{displaymath}
\det (t I + B) \: = \: \sum_{i=0}^n I_{n-i} t^i
\end{displaymath}
For example, $I_0 = 1$, independent of $B$, but the other $I_i$
depend upon $B$.  In particular,
\begin{displaymath}
I_1 \: = \: {\rm tr}\, B,  \: \: \: I_n \: = \: \det B.
\end{displaymath}

In passing, it will sometimes be helpful to define a generalization
of $R_{(r)}$.  For a Young diagram $\mu$, we define $R_{\mu}$ to be
\begin{displaymath}
R_{\mu} \: = \: \det\left[ \begin{array}{ccccc}
R_{(\mu_1)} & R_{(\mu_1+1)} & R_{(\mu_1+2)} & \cdots & R_{(\mu_1+k-1)} \\
\sigma_{(\mu_2-1)} & \sigma_{(\mu_2)} & \sigma_{(\mu_2+1)} & \cdots
&
\sigma_{(\mu_2+k-2)} \\
\sigma_{(\mu_3-2)} & \sigma_{(\mu_3-1)} & \sigma_{(\mu_3)} & \cdots
&
\sigma_{(\mu_3+k-3)} \\
\vdots & & & & \ddots \\
\sigma_{(\mu_k-k+1)} & \sigma_{(\mu_k-k+2)} & \sigma_{(\mu_k-k+3)} &
\cdots & \sigma_{(\mu_k)}
\end{array} \right] .
\end{displaymath}
In the special case that the Young diagram $\mu$ consists of a
single horizontal row of $r$ boxes, which we would label $(r)$, note
that
\begin{displaymath}
R_{\mu} \: = \: R_{(r)},
\end{displaymath}
and it is in this sense that $R_{\mu}$ generalizes $R_{(r)}$.

The description of the ring above holds generically in the space of
tangent bundle deformations, but does break down along certain loci.
Specifically, the description of the classical sheaf cohomology
ring, described by the limit $q \rightarrow 0$, breaks down along
\begin{displaymath}
X \cup V_{n-k+1} \cup V_{n-k+2} \cup \cdots,
\end{displaymath}
where $X$ is the discriminant locus of the tangent bundle
deformation (meaning, the locus where the bundle degenerates), and
$V_m$ is a locus defined by $R_m$, as follows. First, for every
Young diagram $\mu$ of size $|\mu|=m$, such that $\mu_1$, the number
of boxes in the first row, is greater than $n-k$, and no column has
more than $k$ boxes, expand the determinant below in a sum of Schur
polynomials for Young diagrams of the same size:
\begin{displaymath}
R_{\mu} \: = \: \sum_{\nu} C_{\mu \nu}^m \sigma_{\nu},
\end{displaymath}
where $|\nu| = m = |\mu|$. In this fashion, we define a matrix
$(C_{\mu \nu}^m)$. Then, we define $V_m$ to be the locus where the
rank of the (not necessarily square) matrix $(C_{\mu \nu}^m)$ drops.

Along the $V_m$ for
\begin{displaymath}
n-k+1 \: \leq \: m \: \leq \: k(n-k),
\end{displaymath}
the $V_m$ define loci where the dimensions of the sheaf cohomology
groups may jump.  For $m > k(n-k)$, the $V_m$ merely define loci
where the presentation breaks down, where the given relations may
not suffice, but the dimensions of the sheaf cohomology groups do
not jump.

A derivation of this locus is outlined in
appendix~\ref{app:math-rep}. It is useful to note that the locus
above is codimension at least one, and so the presentation of the
classical sheaf cohomology ring is pertinent for generic tangent
bundle deformations.

It is also important to notice that the locus above does not
intersect the (2,2) locus. On the (2,2) locus, where $B=0$ and
$R_{(n)} = \sigma_{(n)}$, from the Giambelli formula $R_{\mu} =
\sigma_{\mu}$, and so we see that along the (2,2) locus, $C^m_{\mu
\nu} = \delta_{\mu\nu}$, whose rank does not drop, and so $V_m$ is
the empty set. Thus, $V_m$ never intersects the (2,2) locus, and
neither does the discriminant.

In passing, note that the result above is consistent with claims of
\cite{ade} that in a sufficiently small neighborhood of the (2,2)
locus, the OPE's can be consistently defined within the topological
subsector.

We conjecture that for $m > k(n-k)$, the loci $V_m$ are all
identical to one another and to the discriminant locus, so that the
total number of components of the locus where the quantum sheaf
cohomology ring relations break down in some fashion is finite.  We
will see this in examples later, though we do not yet have a general
proof for all cases.

So far we have described the loci along which the presentation of
the classical sheaf cohomology ring degenerates.  The degeneration
loci of the presentation of the quantum sheaf cohomology ring are
not completely understood by us at present, though we conjecture
that the same loci $V_m$ are involved, as we shall see in examples
later.

In the remainder of this section, we will check that the ansatz
above correctly specializes to the ordinary classical and quantum
cohomology rings.  We will derive the quantum sheaf cohomology ring
above from the one-loop effective action later in
section~\ref{derivation}.

\subsubsection{Specialization to ordinary classical cohomology}

First, as one extreme, let us reduce to the classical cohomology
ring of $G(k,n)$.  Here, $B=0$ and $q=0$.  In this case,
\begin{displaymath}
R_{(r)} \: = \: \sigma_{(r)},
\end{displaymath}
and the quantum sheaf cohomology ring above becomes
\begin{displaymath}
{\mathbb C}\left[\sigma_{(1)}, \sigma_{(2)}, \cdots  \right] /
\left\langle D_{k+1}, D_{k+2}, \cdots, \sigma_{(n-k+1)},
\sigma_{(n-k+2)}, \cdots
 \right\rangle,
\end{displaymath}
or more simply,
\begin{displaymath}
{\mathbb C}\left[\sigma_{(1)}, \cdots, \sigma_{(n-k)}\right] /
\left\langle D_{k+1}, D_{k+2}, \cdots \right\rangle.
\end{displaymath}
This is almost identical to the presentation of the ordinary
cohomology ring of the Grassmannian given in
equation~(\ref{eq:ord-cohom-grassmannian}), except that here the
relations involve all $D$'s of degree greater than $n$, rather than
going only to $D_n$.  However, we can establish that the two sets of
relations are equivalent, as follows.

We will show that $D_{n+1}$ and all higher relations are linear
combinations of the relations $\{ D_{k+1}, \cdots, D_n\}$, so that
the ring presented above is equivalent to
(\ref{eq:ord-cohom-grassmannian}). To do this, we expand down the
first column of the determinant in the Giambelli formula to derive
the recursion relation
\begin{displaymath}
\sigma_{(\ell,1,\cdots,1)} \: = \: \sigma_{(\ell)} D_{m} \: - \:
\sigma_{(\ell+1,1,\cdots,1)}
\end{displaymath}
where $\sigma_{(\ell,1,\cdots,1)}$ denotes the Schur polynomial
associated to a Young tableau with $\ell$ boxes in the first row and
$1$ box in the next $m$ rows, and $\sigma_{(\ell+1,1,\cdots,1)}$
denotes a similar Schur polynomial, albeit associated to a Young
diagram with $m-1$ rows with one box.  Applying this recursively,
one can quickly show
\begin{equation}  \label{eq:recurse1}
D_{n+1} \: = \: \sigma_{(1)} D_n \: - \: \sigma_{(2)} D_{n-1} \: +
\: \cdots \: + \: (-)^{n-k+1} \sigma_{(n-k)} D_{k+1} \: + \:
(-)^{n-k} \sigma_{(n-k+1,1,\cdots,1)}.
\end{equation}
However, from the Giambelli formula, $\sigma_{(n-k+1,1,\cdots,1)}$
is given by a determinant whose first row vanishes (since it
involves $\sigma$'s all of which are outside the range of the
generators), hence $\sigma_{(n-k+1,1,\cdots,1)} = 0$.  Thus, we see
that $D_{n+1}$ is a linear combination of the relations $\{ D_{k+1},
\cdots, D_n \}$, and one can similarly demonstrate the same result
for all $D_m$ for $m > n$. In this fashion, we see that the ring
above is isomorphic to the ordinary cohomology ring of the
Grassmannian given in equation~(\ref{eq:ord-cohom-grassmannian}).

\subsubsection{Specialization to ordinary quantum cohomology}
\label{sect:check-ord-quant-cohom}

Next, let us verify that the quantum sheaf cohomology
ring~(\ref{eq:general-qsc-ring}) reduces to the ordinary quantum
cohomology ring of the Grassmannian $G(k,n)$ along the (2,2) locus.
This is the case $B=0$, but $q \neq 0$.  As before,
\begin{displaymath}
R_{(r)} \: = \: \sigma_{(r)},
\end{displaymath}
and so the quantum sheaf cohomology ring above becomes
\begin{equation}   \label{eq:ord-q-cohom-bare}
\begin{array}{l}
{\mathbb C}\left[\sigma_{(1)}, \sigma_{(2)}, \cdots \right] /
\left\langle D_{k+1}, D_{k+2}, \cdots,  \sigma_{(n-k+1)}, \cdots,
\sigma_{(n-1)},
\right. \\
\hspace*{1.5in} \left. \sigma_{(n)}+q, \sigma_{(n+1)} + q
\sigma_{(1)}, \cdots
 \right\rangle ,
\end{array}
\end{equation}
specializing for $k=1$ to
\begin{displaymath}
{\mathbb C}\left[\sigma_{(1)}, \sigma_{(2)}, \cdots \right] /
\left\langle D_{2}, D_{3}, \cdots, \sigma_{(n)}+q, \sigma_{(n+1)} +
q \sigma_{(1)}, \cdots
 \right\rangle ,
\end{displaymath}
and for $k=n-1$ to
\begin{displaymath}
{\mathbb C}\left[\sigma_{(1)}, \sigma_{(2)}, \cdots \right] /
\left\langle D_{n}, D_{n+1}, \cdots,  \sigma_{(2)}, \cdots,
\sigma_{(n-1)}, \sigma_{(n)}+q, \sigma_{(n+1)} + q \sigma_{(1)},
\cdots
 \right\rangle .
\end{displaymath}

The expression above for the quantum cohomology ring is not yet in a
standard form, and can be simplified to such a form. First, we show
that, for $\ell \geqslant  1$,
\begin{equation}\label{D(n+l)}
D_{n+\ell} \: = \: 0,
\end{equation}
in the sense that it is redundant, defining no new relations, and
hence can be removed from the presentation above. This can be proved
by induction on $\ell$. First, we will need a small identity. By
expanding the determinant in the definition of $D_m$ across the
first row, (and then expanding the determinant of each submatrix
along the first column), we find
\begin{displaymath}
D_m \: = \: \sigma_{(1)} D_{m-1} - \sigma_{(2)} D_{m-2} \: + \:
\cdots \: + \: (-)^m \sigma_{(m-1)} D_{1} \: + \: (-)^{m+1}
\sigma_{(m)}.
\end{displaymath}
Now, we proceed with the induction. For $\ell=1$, we have
\begin{eqnarray*}
D_{n+1} & = & \sigma_{(1)} D_n + \cdots + (-)^{n-k+1} \sigma_{(n-k)}
D_{k+1} +
(-)^{n-k+2} \sigma_{(n-k+1)} D_{k}\\
& &+\cdots +(-)^n \sigma_{(n-1)} D_2
+ (-)^{n+1} \sigma_{(n)} D_1 + (-)^{n+2} \sigma_{(n+1)}, \\
& = &(-)^{n+1} (\sigma_{(n)} D_1 - \sigma_{(n+1)}), \\
& = &(-)^n q (D_1 - \sigma_{(1)}) \: = \: 0,
\end{eqnarray*}
where we have used the ring relations
\begin{displaymath}
D_{k+1} \: = \: D_{k+2} \: = \: \cdots \: = \: 0, \: \: \:
\sigma_{(n-k+1)} \: = \: \cdots \: = \: \sigma_{(n-1)} \: = \: 0.
\end{displaymath}
Next, assume that \eqref{D(n+l)} is true for all $\ell \leqslant m$.
When $m < k$, we have
\begin{eqnarray*}
D_{n+m+1} & =& \sigma_{(1)} D_{n+m} - \sigma_{(2)} D_{n+m-1} +
\cdots +
(-)^{n+m-k+1} \sigma_{(n+m-k)} D_{k+1} + \\
& & (-)^{n+m-k+2} \sigma_{(n+m-k+1)} D_k + \cdots
+(-)^{n+1} \sigma_{(n)} D_{m+1} + \cdots + (-)^{n+m+2} \sigma_{(n+m+1)},\\
&=& (-)^{n+1} \sigma_{(n)} D_{m+1} + \cdots + (-)^{n+m+2} \sigma_{(n+m+1)},\\
&=& (-)^{n} q \left(D_{m+1} - \sigma_{(1)} D_{m} + \cdots +
(-)^{m+1} \sigma_{(m+1)}\right) \: = \: 0.
\end{eqnarray*}
When $m \geqslant k$, we have
\begin{eqnarray*}
D_{n+m+1} &=& \sigma_{(1)} D_{n+m} - \sigma_{(2)} D_{n+m-1} + \cdots
+
(-)^{n} \sigma_{(n-1)} D_{m+2}  \\
& & + (-)^{n+1} \sigma_{(n)} D_{m+1} + \cdots
 + (-)^{n+m+2} \sigma_{(n+m+1)}, \\
& =& (-)^{n+1} \sigma_{(n)} D_{m+1} + \cdots + (-)^{n+m+2} \sigma_{(n+m+1)},\\
&=& (-)^{n} q (D_{m+1} - \sigma_{(1)} D_{m} + \cdots + (-)^{m+1}
\sigma_{(m+1)}) \: = \: 0.
\end{eqnarray*}
Thus, we have shown~(\ref{D(n+l)}).

Next, using the relations
\begin{displaymath}
\sigma_{(n+\ell)} \: = \: - q \sigma_{(\ell)},
\end{displaymath}
we can express the $\sigma_{(i)}$'s with $i > n-k$ as polynomials of
$q$ and $\sigma_{(i)}$'s with $i \leqslant n-k$, and so we can
rewrite the ring in terms of generators
\begin{displaymath}
\sigma_{(1)}, \cdots, \sigma_{(n-k)}.
\end{displaymath}

Finally, we derive an expression for $D_n$. Starting with
\begin{eqnarray*}
D_n & = & \sigma_{(1)} D_{n-1} - \sigma_{(2)} D_{n-2} + \cdots +
(-)^{n-k} \sigma_{(n-k-1)} D_{k+1} + (-)^{n-k-1} \sigma_{(n-k)} D_{k} \\
& & (-)^{n-k+2} \sigma_{(n-k+1)} D_{k-1} + \cdots + (-)^n
\sigma_{(n-1)} D_1 + (-)^{n+1} \sigma_{(n)},
\end{eqnarray*}
we use the ring relations
\begin{displaymath}
D_{k+1} \: = \: \cdots \: = \: D_{k(n-k)} \: = \: 0, \: \: \:
\sigma_{(n-k+1)} \: = \: \cdots \: = \: \sigma_{(n-1)} \: = \: 0
\end{displaymath}
and the fact that
\begin{displaymath}
n-1 \: \leq \: k(n-k)
\end{displaymath}
to simplify $D_n$ to
\begin{displaymath}
D_n \: = \: (-)^{n-k-1} \sigma_{(n-k)} D_{k} + (-)^{n+1}
\sigma_{(n)},
\end{displaymath}
which using further ring relations can be written as
\begin{displaymath}
D_n \: = \:  (-)^{n-k-1} \sigma_{(n-k)} D_{k} + (-)^n q .
\end{displaymath}

Finally, let us simplify the ring
presentation~(\ref{eq:ord-q-cohom-bare}). We have shown that inside
that quotient ring, $D_i$ is redundant for $i>n$, and given our
expression for $D_n$ above, it is straightforward to see that the
ring~(\ref{eq:ord-q-cohom-bare}) can be reduced to
\begin{equation}
\mathbb{C}\left[\sigma_{(1)},\cdots,\sigma_{(n-k)}\right]/
\left\langle D_{k+1},\cdots,D_{n-1}, D_n+(-)^n q \right\rangle,
\end{equation}
where the ``$D_n$'' above is the `classical' $D_n$, namely
\begin{displaymath}
(-)^{n-k-1} \sigma_{(n-k)} D_{k}
\end{displaymath}
in the notation of~(\ref{eq:ord-q-cohom-bare}).  This new
presentation is a standard representation of the quantum cohomology
ring of $G(k,n)$ (see {\it e.g.}
\cite{buch1,buch2,tamvakisrev,bertram1,edver,st}).
\par

\subsection{Derivation from one-loop effective action}\label{derivation}

In this section we will describe how the quantum sheaf cohomology
ring relations can be computed from the one-loop effective action,
and in the next section we will check our results against A/2
correlation functions computed via supersymmetric localization. (The
classical sheaf cohomology ring relations will be derived
mathematically in the companion paper~\cite{3}; a purely
mathematical derivation of the quantum sheaf cohomology ring
relations here is left for future work.)

Before computing quantum sheaf cohomology for general bundle
deformations, we shall begin by deriving the ordinary quantum
cohomology, along the (2,2) locus, from the one-loop effective
action, as a warm-up exercise.

As before, let us take the diagonal elements of the $\sigma$ field
to be $\sigma_i, i=1,\cdots,k$.  On the (2,2) locus, where $B=0$, we
see from the one-loop effective potential \eqref{J(0,2)} that the
$\sigma_i$ obey equation~(\ref{emo}), or more simply
\[
\sigma_i^n = -q, \: \: \: i=1,\cdots,k.
\]
Because these equations are of order $n$, all relations with order
lower than $n$ are not affected, {\it i.e.}
\[
\sigma_{(i)}=0,~ \: \: i=n-k+1,\cdots,n-1.
\]
Then, for example, from
\[
\sigma_{(n-1)} \sigma_1 = \sum_{\alpha_1+\cdots+\alpha_k = n \atop
\alpha_1 \neq 0} \sigma_1^{\alpha_1} \cdots \sigma_k^{\alpha_k}=0,
\]
we get
\[
\sigma_{(n)} = \sum_{\alpha_1+\cdots+\alpha_k = n}
\sigma_1^{\alpha_1} \cdots \sigma_k^{\alpha_k} =
\sum_{\alpha_2+\cdots+\alpha_k = n} \sigma_2^{\alpha_2} \cdots
\sigma_k^{\alpha_k}.
\]
Similarly, from $\sigma_{(n-2)} \sigma_1=0$, we have
\[
\sigma_{(n-1)} = \sum_{\alpha_2+\cdots+\alpha_k = n-1}
\sigma_2^{\alpha_2} \cdots \sigma_k^{\alpha_k}.
\]
Then $\sigma_{(n-1)} \sigma_2 = 0$ shows that
\[
\sigma_{(n)} = \sum_{\alpha_3+\cdots+\alpha_k = n}
\sigma_3^{\alpha_3} \cdots \sigma_k^{\alpha_k}.
\]
This procedure stops in $k-1$ steps, and we get our first quantum
corrected relation
\begin{equation}   \label{eq:22-step1}
\sigma_{(n)} = \sigma_k^n = -q.
\end{equation}
To derive the equation above, we arbitrarily made use of $\sigma_1,
\cdots, \sigma_{k-1}$; by picking a different set of $k-1$
$\sigma_i$'s, we arrive at equation~(\ref{eq:22-step1}) for each
value of $k$.

We can use the same method to deduce the higher order relations, for
example
\[
\sigma_{(n)} \sigma_1 = -q \sigma_1 = \sum_{\alpha_1+\cdots+\alpha_k
= n+1 \atop \alpha_1 \neq 0} \sigma_1^{\alpha_1} \cdots
\sigma_k^{\alpha_k},
\]
which implies
\[
\sigma_{(n+1)} = \left( \sum_{\alpha_2+\cdots+\alpha_k = n+1}
\sigma_2^{\alpha_2} \cdots \sigma_k^{\alpha_k}\right)  -q \sigma_1.
\]
Repeatedly using lower order relations leads to
\[
\sigma_{(n+1)} \: =\:  -q \sigma_1 - q \sigma_2 -\cdots - q \sigma_k
\: = \: -q \sigma_{(1)}.
\]
Similarly, one can compute
\[
\sigma_{(n+l)} = -q \sigma_{(l)},~ l\geqslant 1.
\]

In this fashion, we find that the one-loop effective action implies
a quantum cohomology ring of the form~(\ref{eq:ord-q-cohom-bare}),
which we show in section~\ref{sect:check-ord-quant-cohom} matches
standard presentations of the ordinary quantum cohomology ring.

So far we have shown how one-loop effective action arguments can be
used to derive the ordinary quantum cohomology ring along the (2,2)
locus. Next we shall leave the (2,2) locus and consider more general
(0,2) cases by turning on a nonzero $B$ deformation.

First, we note that the relations
\begin{displaymath}
D_{k+1} \: = \: D_{k+2} \: = \: \cdots \: = \: 0
\end{displaymath}
are trivially satisfied for all $\sigma_{(k)}$ constructed as Schur
polynomials in $k$ variables $\sigma_1, \cdots, \sigma_k$. It
remains to derive the relations
\begin{displaymath}
R_{(n-k+1)} \: = \: \cdots \: = \: R_{(n-1)} \: = \: 0, \: \: \:
R_{(n)} \: = \: -q, \: \: \: R_{(n+1)} \: = \: -q \sigma_{(1)}, \:
\: \: \cdots
\end{displaymath}

For any $n \times n$ matrix $B$, the quantum corrected relations are
encoded in
\begin{equation}\label{detE=-q}
\det(E(\sigma_\alpha)) = \det(\sigma_\alpha I + B \sigma_{(1)}) = -q
\end{equation}
due to the one-loop effective potential \eqref{J(0,2)}. Note that,
by definition of $I_i$ we have
\begin{equation}\label{detE exp}
\det(E(\sigma_a)) = \sum_{i=0}^n I_i \sigma_{(1)}^i
\sigma_{a}^{n-i}.
\end{equation}
Again, the relations with dimension smaller than $n$ do not receive
quantum corrections, {\it i.e.} the relations
\begin{equation}\label{lower relations}
R_{(n-k+1)}=R_{(n-k+2)}=\cdots=R_{(n-1)}=0
\end{equation}
still hold in the quantum case. Now let's compute the relation at
order $n$. First, note
\[
R_{(n-1)} \sigma_1 = \sum_{i=0}^{n-1} I_i \sigma_{(n-i-1)}
\sigma_{(1)}^i \sigma_1 =\sum_{i=0}^{n-1} I_i \left(\sigma_{(n-i)}-
\sum_{|\alpha|=n-i \atop \alpha_1=0} \sigma^{[\alpha]}\right)
\sigma_{(1)}^i = 0,
\]
where $\sigma^{[\alpha]}$ denotes $\sigma_1^{\alpha_1}
\sigma_2^{\alpha_2} \cdots \sigma_k^{\alpha_k}$, $\alpha$ is the
corresponding multi-index, and we have used the relation
$R_{(n-1)}=0$. This implies that
\[
R_{(n)}=\sum_{i=0}^{n} I_i \sigma_{(n-i)} \sigma_{(1)}^i =
\sum_{i=0}^{n} I_i \left( \sum_{|\alpha|=n-i \atop \alpha_1=0}
\sigma^{[\alpha]}\right) \sigma_{(1)}^i .
\]
Similarly, $R_{(n-2)} \sigma_1 = 0$ implies
\[
R_{(n-1)} = \sum_{i=0}^{n-1} I_i \left( \sum_{|\alpha|=n-i-1 \atop
\alpha_1=0} \sigma^{[\alpha]}\right) \sigma_{(1)}^i,
\]
and $R_{(n-1)} \sigma_2 = 0$ leads to
\[
R_{(n)} = \sum_{i=0}^{n} I_i \left( \sum_{|\alpha|=n-i \atop
\alpha_1=\alpha_2=0} \sigma^{[\alpha]}\right) \sigma_{(1)}^i.
\]
Because we have $k-1$ relations in \eqref{lower relations},
induction shows that this procedure allows us to eliminate
$\sigma_1$ through $\sigma_{k-1}$ in the expression of $R_n$, {\it
i.e.}
\[
R_{(n)}=\sum_{i=0}^{n} I_i \sigma_k^{n-i} \sigma_{(1)}^i,
\]
which is equal to $-q$ due to \eqref{detE=-q} and \eqref{detE exp}.
Thus the relation at order $n$ is
\begin{equation}\label{Rn}
R_{(n)}+q=0.
\end{equation}
We can follow the same procedure to compute the quantum correction
to $R_{(n+1)}$. Since now there are $k$ relations at hand including
\eqref{Rn}, all the $\sigma_i$ dependence can be eliminated except
those proportional to $q$. One can compute
\begin{equation}
R_{(n+1)} + q \sigma_{(1)} = 0.
\end{equation}
In general, we can show
\begin{equation}\label{general R_(n+l)}
R_{(n+\ell)} = -q \sigma_{(\ell-1)} P_1 + q \sigma_{(\ell-2)} P_2 -
q \sigma_{(\ell-3)} P_3 + \cdots + (-)^k q \sigma_{(\ell-k)} P_k,
\end{equation}
where $\ell>0$ and $P_i$ is the elementary symmetric polynomial of
order $i$ in $\sigma_1,\cdots, \sigma_k$. Again, we define
$\sigma_{(s)}=0$ when $s<0$ in the above formula. From the fact that
\[
\begin{split}
\prod_{i=1}^k (1+\sigma_i t)^{-1} &= \sum_{j=0}^\infty (-)^j \sigma_{(j)} t^j \\
\prod_{i=1}^k (1+\sigma_i t) &= \sum_{j=0}^k P_j t^j,
\end{split}
\]
we have
\[
\sigma_{(\ell)}-\sigma_{(\ell-1)} P_1 + \sigma_{(\ell-2)} P_2 +
\cdots + (-)^k \sigma_{(\ell-k)} P_k = 0,
\]
which implies
\begin{equation}\label{general R}
R_{(n+\ell)}+q \sigma_{(\ell)} = 0,\: \: \: ~\ell \geqslant 0.
\end{equation}
Actually, \eqref{general R}, or equivalently \eqref{general
R_(n+l)}, can be proved by induction with the same method as for
$R_{(n)}$ and $R_{(n+1)}$. Indeed, if we assume \eqref{general R} is
true for $l$ replaced with any positive integer smaller than $l$, we
get
\begin{displaymath}
R_{(n+\ell-1)} \sigma_1 \: = \: \sum_{i=0}^n I_i \left(
\sum_{|\alpha|=n+\ell-i \atop \alpha_1 \neq =0}
\sigma^{[\alpha]}\right) \sigma_{(1)}^i ,
\end{displaymath}
must match
\begin{displaymath}
-q \sigma_{(\ell-1)} \sigma_1,
\end{displaymath}
(by the inductive assumption) and hence, if we define $E_{s,t}$ to
be the elementary polynomial of order $t$ in $\sigma_1,\cdots,
\sigma_s$, for $0\leqslant s \leqslant k$ and $t \leqslant s$, then
\[
\begin{split}
R_{(n+\ell)}= \sum_{i=0}^n I_i \left( \sum_{|\alpha|=n+\ell-i \atop
\alpha_1=0} \sigma^{[\alpha]}\right) \sigma_{(1)}^i - q
\sigma_{(\ell-1)} E_{1,1} .
\end{split}
\]
Let's suppose that
\begin{equation}\label{induction}
\begin{split}
R_{(n+\ell-s)}=&\sum_{i=0}^n I_i \left( \sum_{|\alpha|=n+\ell-s-i
\atop \alpha_1=\cdots = \alpha_{u-s}=0} \sigma^{[\alpha]} \right)
\sigma_{(1)}^i -q \sigma_{(\ell-s-1)} E_{u-s,1}+q
\sigma_{(\ell-s-2)} E_{u-s,2}+ \\
 &\cdots +(-)^{u-s}
q \sigma_{(\ell-t)} E_{u-s,u-s}
\end{split}
\end{equation}
for any $u \leqslant t<k$ and $0\leqslant s \leqslant u$ (we have
seen this is true for $t=1$). Starting with
\[
R_{(n+\ell-t)} = \sum_{i=0}^n I_i \left( \sum_{|\alpha|=n+\ell-t-i
\atop \alpha_1=0} \sigma^{[\alpha]}\right) \sigma_{(1)}^i - q
\sigma_{(\ell-t-1)} \sigma_1,
\]
which is obtained from $R_{(n+\ell-t-1)} \sigma_1 = -q
\sigma_{(\ell-t-1)} \sigma_1$, induction on $s$ shows
\eqref{induction} is valid for $u \leqslant t+1$ and $s \leqslant
u$. Thus we can take $u=k$ and $s=0$ in \eqref{induction} to get
\[
R_{(n+\ell)} =
 -q \sigma_{(\ell-1)} E_{k,1}+q \sigma_{(\ell-2)} E_{k,2}+\cdots +
(-)^k q  \sigma_{(\ell-k)} E_{k,k},
\]
which is exactly \eqref{general R_(n+l)}, hence
proving~(\ref{general R}).

\section{Examples}\label{examples}

In this section, we will perform consistency tests on the quantum
sheaf cohomolgy ring~(\ref{eq:general-qsc-ring}) by using
supersymmetric localization to compute A/2 correlation functions in
examples, and check that the predictions of the quantum sheaf
cohomology ring are consistent with those correlation functions.

In each example, we will begin by describing correlation functions
and quantum cohomology along the (2,2) locus, and will generalize to
(0,2). Furthermore, in all our (0,2) examples, we will take $B$ to
be diagonal:
\begin{displaymath}
B \: = \: {\rm diag}(b_1, \cdots, b_n)
\end{displaymath}
on $G(k,n)$.  The methods of this paper apply to general $B$;
however, the resulting formulas for general $B$ are rather complex,
and it suffices to consider the special case of $B$ diagonal for the
purposes of illustrative examples.

We will begin by looking at examples of projective spaces as special
cases of the construction described here, and then will turn to
Grassmannians which are not projective spaces.

\subsection{$G(1,3)$}

The Grassmannian $G(1,3)$ is the projective space ${\mathbb P}^2$,
so any tangent bundle deformation is trivial -- the tangent bundle
is rigid.  Nevertheless, this example and $G(2,3)$ will serve as
simple prototypes for later results.

Let us begin by computing correlation functions on the (2,2) locus.
Since $G(1,3)$ is described by a $U(1)$ gauge theory, there is only
a single $\sigma$ field. Here, the localization
formula~(\ref{localization(0,2)}) implies that classical (two-point)
correlation functions are given by
\begin{displaymath}
\langle f(\sigma) \rangle \: = \: {\rm JKG-Res}\left\{
\frac{1}{\sigma_1^3} f(\sigma) \right\},
\end{displaymath}
which trivially reduces to the ordinary one-dimensional residue. The
only nonvanishing classical correlation function is given by
\begin{displaymath}
\langle \sigma^2 \rangle \: = \: 1 \: = \: \langle
\sigma_{\tiny\yng(2)} \rangle
\end{displaymath}
Similarly, the one-instanton contributions to correlation functions
are given by
\begin{displaymath}
\langle f(\sigma) \rangle \: = \: {\rm JKG-Res}\left\{
\frac{q}{\sigma_1^6} f(\sigma) \right\},
\end{displaymath}
which is again an ordinary contour integral about $\sigma=0$.
Clearly the only nonvanishing correlation function in the
one-instanton sector is
\begin{displaymath}
\langle \sigma^5 \rangle \: = \: q,
\end{displaymath}
and so we read off the quantum cohomology relation
\begin{displaymath}
\sigma^3 \: = \: \sigma_{\tiny\yng(3)} \: = \: q.
\end{displaymath}
This reproduces the quantum cohomology ring of ${\mathbb P}^2$,
given by
\begin{displaymath}
{\mathbb C}[x]/(x^3-q).
\end{displaymath}

After the (trivial) (0,2) deformation, classical correlation
functions take the form
\begin{displaymath}
\langle f(\sigma) \rangle \: = \: {\rm JKG-Res}\left\{
\left(\frac{1}{\det \tilde{E}(\sigma)} \right) f(\sigma) \right\} ,
\end{displaymath}
where the JKG residue is trivially an ordinary residue, and
\begin{displaymath}
\tilde{E}(\sigma) \: = \: (I + B)(\sigma)
\end{displaymath}
for the case we shall consider, hence
\begin{displaymath}
\det \tilde{E}(\sigma) \: = \: \sigma^3 \det(I+B).
\end{displaymath}
The only nonzero classical correlation function is
\begin{displaymath}
\langle \sigma^2 \rangle \: = \: \langle \sigma_{\tiny\yng(2)}
\rangle \: = \: \frac{1}{\det(I+B)}.
\end{displaymath}
Similarly, in the one-instanton sector,
\begin{displaymath}
\langle f(\sigma) \rangle \: = \: {\rm JKG-Res}\left\{ q
\left(\frac{1}{\det \tilde{E}(\sigma)} \right)^2 f(\sigma) \right\}
,
\end{displaymath}
where again the JKG residue is an ordinary residue at $\sigma=0$,
and the only nontrivial correlation function is
\begin{displaymath}
\langle \sigma^5 \rangle \: = \: \frac{1}{( \det(I+B) )^2}.
\end{displaymath}

From the structure of these correlation functions, as well as the
one-loop effective action, one can read off that the quantum ring
relation is given by
\begin{displaymath}
\det \tilde{E}(\sigma) \: = \: q
\end{displaymath}
or equivalently
\begin{displaymath}
\sigma^3 \: = \: \sigma_{\tiny\yng(3)} \det(I+B) \: = \: q.
\end{displaymath}
By a simple redefinition of $q$, the resulting ring is identical to
that on the (2,2) locus, as expected for a trivial bundle
deformation.

Now, let us compare to the prediction of the quantum sheaf
cohomology ring~(\ref{eq:general-qsc-ring}).  In this case, the ring
should be given by
\begin{displaymath}
{\mathbb C}\left[\sigma_{(1)}, \sigma_{(2)},\cdots \right] /
\left\langle D_2, \cdots, R_{(3)}+q, R_{(4)} + q \sigma_{(1)},
\cdots \right\rangle \: = \: {\mathbb C}\left[\sigma_{(1)}\right] /
\left\langle R_{(3)}+q \right\rangle.
\end{displaymath}
In writing the above, we have used the fact that
\begin{displaymath}
D_2 \: = \: \sigma_{(1)}^2 - \sigma_{(2)},
\end{displaymath}
hence $\sigma_{(2)}$ (and, similarly, higher $\sigma_{(i)}$) are
redundant, and also the consequence
\begin{displaymath}
R_{(3 + \ell)} + q \sigma_{(\ell)} \: = \: \sigma_{(\ell)} \left(
R_{(3)} + q \right),
\end{displaymath}
which makes the higher $R_{(n)}$ relations redundant. Finally, note
that
\begin{eqnarray*}
R_{(3)} & = & \sum_{i=0}^3 I_i \sigma_{(3-i)} \sigma^i , \\
& = & \left( \sum_{i=0}^3 I_i \right) \sigma^3, \\
& = & \left( \det(I+B) \right) \sigma^3,
\end{eqnarray*}
Clearly, this ring precisely matches the relation derived above.

Let us conclude by computing the locus on which the sheaf cohomology
jumps, defined by
\begin{displaymath}
V_{3} \cup V_4 \cup V_5 \cup \cdots
\end{displaymath}
as described in section~\ref{sect:qsc-ring-details}. Since $G(1,3)$
is a projective space, and projective spaces admit no tangent bundle
deformations, we should recover no more than the discriminant locus,
but this will be both a good consistency test as well as an explicit
demonstration of the $V$'s.

We begin by computing $V_{2+\ell}$. There is only one Young diagram
with $2+\ell$ boxes, at least two along the first row, and none in
later rows, namely $(2+\ell)$, and trivially
\begin{displaymath}
R_{(2+\ell)} \: = \: (1 + I_1 + I_2 + I_3) \sigma_{(2+\ell)} \: = \:
\sigma_{(2+\ell)} \det(1+B).
\end{displaymath}
Thus, for any $V_{2+\ell}$, the matrix $C_{\mu \nu}$ is a $1 \times
1$ matrix, with single component
\begin{displaymath}
\det(1+B),
\end{displaymath}
and so we see that for all $\ell \geq 1$, the locus $V_{2+\ell}$
coincides with the discriminant locus.  Thus, there are no new
components, as expected from the fact that (for $B$ not in the
discriminant locus) the tangent bundle deformations are all trivial.

It is easy to see that a nearly identical argument holds for
$G(1,n)$, that all of the loci $V_{n-k+\ell}$ for such Grassmannians
are copies of the discriminant locus.

\subsection{$G(2,3)$}

Before moving on to Grassmannians with nontrivial bundle
deformations, let us look at a different presentation of ${\mathbb
P}^2$, as a $U(2)$ gauge theory rather than a $U(1)$ gauge theory.

As before, let us first examine the (2,2) locus. In this theory, the
classical (two-point) functions are given by
\begin{displaymath}
\langle f(\sigma) \rangle \: = \: \frac{1}{2!} {\rm JKG-Res}\,
\left\{ (-) (\sigma_1-\sigma_2)^2 \frac{1}{\sigma_1^3}
\frac{1}{\sigma_2^3} f(\sigma) \right\},
\end{displaymath}
or explicitly,
\begin{eqnarray*}
\langle \sigma_1^2 \rangle & = & -\frac{1}{2!}, \\
\langle \sigma_1 \sigma_2 \rangle & = & + \frac{2}{2!}, \\
\langle \sigma_2^2 \rangle & = & - \frac{1}{2!},
\end{eqnarray*}
using the fact that the JKG residue in this case is just iterated
ordinary residues at $\sigma_2=0$ and $\sigma_1=0$. We interpret the
$\sigma_i$'s on the (2,2) locus as Chern roots of the universal
subbundle $S$.  In that language, the degree two cohomology is
generated by
\begin{displaymath}
\sigma_{\tiny\yng(1)} \: = \: \sigma_1 + \sigma_2,
\end{displaymath}
corresponding to a (1,1) form on ${\mathbb P}^2$, and the degree
four cohomology is generated by
\begin{displaymath}
\sigma_{\tiny\yng(1,1)} \: = \: \sigma_1 \sigma_2 ,
\end{displaymath}
corresponding to a (2,2) form on ${\mathbb P}^2$. As a consistency
check, note that
\begin{displaymath}
\langle \sigma_{\tiny\yng(1,1)} \rangle \: = \: \langle \sigma_1
\sigma_2 \rangle \: \neq \: 0,
\end{displaymath}
as expected since $\sigma_{\tiny\yng(1,1)}$ should correspond to a
top-form.

Furthermore, as all the contributing Schubert cells are associated
with subdiagrams of the $k\times(n-k)$ box
\begin{displaymath}
\yng(1,1)
\end{displaymath}
there is a relation
\begin{displaymath}
\sigma_{\tiny\yng(2)} \: = \: 0
\end{displaymath}
which we can check explicitly.  In terms of Schur polynomials,
\begin{displaymath}
\sigma_{\tiny\yng(2)} \: = \: \sigma_1^2 \: + \: \sigma_2^2 \: + \:
\sigma_1 \sigma_2
\end{displaymath}
and it is immediate that
\begin{displaymath}
\langle \sigma_1^2 \rangle \: + \: \langle \sigma_1 \sigma_2 \rangle
\: + \: \langle \sigma_2^2 \rangle \: = \: 0
\end{displaymath}

Note in passing that with the relation $\sigma_{\tiny\yng(2)} = 0$,
we have that $\sigma_{\tiny\yng(1)}^2 = \sigma_{\tiny\yng(1,1)}$, as
expected for the cohomology ring of ${\mathbb P}^2$.

Thus, in this fashion we can interpret the $\sigma_{1,2}$ and see
the cohomology of $G(2,3) = {\mathbb P}^2$ in the correlation
functions above.

Next, let us turn to the formal (0,2) deformations of this theory.
Now, ultimately because the tangent bundle of ${\mathbb P}^2$ has no
nontrivial deformations, we should get isomorphic results, but this
is a good warm-up exercise for nontrivial examples later.

Here, classical correlation functions have the form
\begin{displaymath}
\langle f(\sigma) \rangle \: = \: \frac{1}{2!} {\rm JKG-Res}\left\{
(-) (\sigma_1-\sigma_2)^2 \left( \frac{1}{\det \tilde{E}(\sigma_1)}
\right) \left( \frac{1}{\det \tilde{E}(\sigma_2)} \right) f(\sigma)
\right\}
\end{displaymath}
where
\begin{displaymath}
\tilde{E}(x) \: = \: I x \: + \: B (\sigma_1 + \sigma_2).
\end{displaymath}
The JKG residue above is computed as iterated ordinary residues, by
first summing the residues about
\begin{displaymath}
\sigma_1 \: = \: - \sigma_2 \frac{b_1}{1+b_1}, \: \: \: - \sigma_2
\frac{b_2}{1+b_2}, \: \: \: - \sigma_2 \frac{b_3}{1+b_3},
\end{displaymath}
corresponding to the zeroes of $\det \tilde{E}(\sigma_1)$, and then
taking the residue about $\sigma_2=0$.

Define
\begin{eqnarray*}
\Delta & = & 2 \prod_{i<j} (1 + b_{i} + b_{j} ), \\
& = & 2\left( 1 + 2 I_1 + I_1^2 + I_2 + I_1 I_2 - I_3 \right),
\end{eqnarray*}
then
\begin{eqnarray*}
\langle \sigma_1^2 \rangle & = & \Delta^{-1}\left[ -1 -2 I_2 -2 I_1
\right] , \\
& & \\
\langle \sigma_1 \sigma_2 \rangle & = & \Delta^{-1} \left[ 2 + 2 I_2
+ 2 I_1
\right] , \\
& & \\
\langle \sigma_2^2 \rangle & = & \Delta^{-1} \left[ -1 -2 I_2 -2 I_1
\right] \: = \: \langle \sigma_1^2 \rangle ,
\end{eqnarray*}
or equivalently
\begin{eqnarray*}
\langle \sigma_{\tiny\yng(2)} \rangle & = & \langle \sigma_1^2
\rangle \: + \: \langle \sigma_1 \sigma_2 \rangle \: + \:
\langle \sigma_2^2 \rangle , \\
& = & \Delta^{-1} \left[ -2 I_2 -2 I_1
\right] , \\
& & \\
\langle \sigma_{\tiny\yng(1)}^2 \rangle & = & \langle \sigma_1^2
\rangle \: + \: 2 \langle \sigma_1 \sigma_2 \rangle
\: + \: \langle \sigma_2^2 \rangle, \\
& = & 2 \Delta^{-1} , \\
& & \\
\langle \sigma_{\tiny\yng(1,1) } \rangle & = &
\langle \sigma_1 \sigma_2 \rangle , \\
& = & \Delta^{-1} \left[ 2 + 2 I_2 + 2 I_1 \right] ,
\end{eqnarray*}
where
\begin{eqnarray*}
I_3 & = & b_1 b_2 b_3 \: = \: \det B, \\
I_2 & = & \sum_{i<j} b_{i} b_{j} , \\
I_1 & = & \sum_i b_{i} \: = \: {\rm tr}\, B.
\end{eqnarray*}

Now, the quantum sheaf cohomology ring predicted
by~(\ref{eq:general-qsc-ring}) is of the form
\begin{displaymath}
{\mathbb C}\left[\sigma_{(1)},\sigma_{(2)}, \cdots \right] /
\left\langle D_3, D_4, \cdots, R_{(2)}, R_{(3)} + q, R_{(4)} + q
\sigma_{(1)}, \cdots \right\rangle.
\end{displaymath}
The relations
\begin{displaymath}
R_{(2)} \: = \: D_3 \: = \: D_4 \: = \:\cdots \: = \: 0
\end{displaymath}
allow one to write $\sigma_{(2)}$ and all higher $\sigma_{(r)}$ as
linear combinations of powers of $\sigma_{(1)}$, so that there is,
in effect, only one generator.  As the tangent bundle of $G(2,3) =
{\mathbb P}^2$ is rigid, the quantum sheaf cohomology ring should
be, for nondegenerate cases, isomorphic to the ordinary quantum
cohomology ring of ${\mathbb P}^2$, so indeed only one generator is
expected.

Now,
\begin{eqnarray*}
R_{(2)} & = & \sum_{i=0}^2 I_i \sigma_{(2-i)} \sigma_{(1)}^i, \\
& = & \sigma_{(2)} + (I_1 + I_2) \sigma_{(1)}^2,
\end{eqnarray*}
and
\begin{eqnarray*}
R_{(3)} & = & \sum_{i=0}^3 I_i \sigma_{(3-i)} \sigma_{(1)}^i, \\
& = & \sigma_{(3)} + I_1 \sigma_{(2)} \sigma_{(1)} + (I_2 + I_3)
\sigma_{(1)}^3.
\end{eqnarray*}

Note that, with a bit of algebra, the relations
\begin{displaymath}
D_3 \: = \: 0 \: = \: R_{(2)} \: = \: R_{(3)}+q
\end{displaymath}
can be solved for $\sigma_{(2)}$ and $\sigma_{(3)}$ to derive the
relation
\begin{equation} \label{eq:ex:g23:1st-gen}
\left( 1 + 2 I_1 + I_2 - I_3 + I_1(I_1 + I_2) \right) \sigma_{(1)}^3
\: = \:
 - q.
\end{equation}
Thus, for generic tangent bundle degenerations, we recover the
ordinary quantum cohomology ring relation for ${\mathbb P}^2$, up to
an irrelevant scaling.  The coefficient of $\sigma_{(1)}^3$ vanishes
on the discriminant locus $\{\Delta=0\}$ (which matches the $V_3$
locus, as we shall see later).

We can see the ring relations in the correlation functions above as
follows. First, note that already in the classical correlation
functions,
\begin{displaymath}
\langle R_{(2)} \rangle \: = \: \langle \sigma_{\tiny\yng(2)}
\rangle + (I_1 + I_2) \langle \sigma_{\tiny\yng(1)}^2 \rangle
 \: = \: \Delta^{-1} \left[
-2 (I_1+I_2) + (I_1 + I_2)(2) \right] \: = \: 0,
\end{displaymath}
as one would expect from the ring relations above. On the (2,2)
locus, $R_{(2)}$ specializes to $\sigma_{\tiny\yng(2)}$, and so this
becomes the relation $\sigma_{\tiny\yng(2)}=0$ discussed earlier.

Now, let us compute the $V$ loci described in
section~\ref{sect:qsc-ring-details}, where the dimensions of the
classical sheaf cohomology groups jump and our description of the
quantum sheaf cohomology ring breaks down.  As $G(2,3)$ is a
projective space, and projective spaces admit no nontrivial tangent
bundle deformations, we should find that the $V$ loci contain
nothing more than the discriminant locus.

First, we compute $V_2$.  For this, there is only one Young diagram
with two boxes total and more than one box in the first row, namely
$(2)$.  We write
\begin{displaymath}
R_{(2)} \: = \: R_{\tiny\yng(2)} \: = \: \sigma_{(2)} + (I_1 + I_2)
\sigma_{(1)}^2 \: = \: \sigma_{(2)} \left( 1 + I_1 + I_2 \right) \:
+ \: \sigma_{(1,1)}\left( I_1 + I_2 \right).
\end{displaymath}
Define the matrix
\begin{displaymath}
\left( C^2_{\mu \nu} \right) \: = \: \left[ 1 + I_1 + I_2, I_1 + I_2
\right],
\end{displaymath}
so that
\begin{displaymath}
\left[ R_{(2)} \right] \: = \: \left( C^2_{\mu \nu} \right) \left[
\begin{array}{c}
\sigma_{(2)} \\
\sigma_{(1,1)} \end{array} \right],
\end{displaymath}
then the locus $V_2$ is defined to be the locus on which $(C^2_{\mu
\nu})$ drops rank.  However, that would require
\begin{displaymath}
1 + I_1 + I_2 \: = \: 0 \: = \: I_1 + I_2,
\end{displaymath}
which has no solutions, therefore $V_2$ is the empty set.

Next, we compute $V_3$. Here, there are two Young diagrams with
three boxes total and more than one box in the first row, namely
$(3)$ and $(2,1)$.  We compute
\begin{eqnarray*}
R_{(3)} \: = \: R_{\tiny\yng(3)} & = & \sigma_{(3)} + I_1
\sigma_{(2)} \sigma_{(1)}
+ (I_2 + I_3) \sigma_{(1)}^3 , \\
& = & \sigma_{(3)} \left( 1 + I_1 + I_2 + I_3 \right) \: + \:
\sigma_{(2,1)} \left( I_1 + 2 I_2 + 2 I_3 \right),
\end{eqnarray*}
and
\begin{eqnarray*}
R_{(2,1)} \: = \: R_{\tiny\yng(2,1)} & = & \det\left[
\begin{array}{cc}
R_{(2)} & R_{(3)} \\
1 & \sigma_{(1)} \end{array} \right] \: = \:
\sigma_{(1)} R_{(2)} - R_{(3)}, \\
& = & \sigma_{(3)}\left( - I_3 \right) \: + \: \sigma_{(2,1)} \left(
1 + I_1 - 2 I_3 \right).
\end{eqnarray*}
Thus, the matrix $(C^3_{\mu \nu})$ is given by
\begin{displaymath}
\left( C^3_{\mu \nu} \right) \: = \: \left[ \begin{array}{cc}
1 + I_1 + I_2 + I_3 & I_1 + 2I_2 + 2 I_3  \\
-I_3 & 1+I_1 -2I_3
\end{array} \right],
\end{displaymath},
and the locus $V_3$ is then defined as
\begin{displaymath}
V_3 \: = \: \{ \det (C^3_{\mu \nu}) = 0 \}.
\end{displaymath}
It is straightforward to check that this locus matches the
discriminant locus $X = \{\Delta=0\}$.

With a little algebra, it is straightforward to verify that the ring
relation~(\ref{eq:ex:g23:1st-gen}) can be rewritten as
\begin{displaymath}
(\det C^3_{\mu \nu}) \sigma_{(1)}^3 \: = \: -q,
\end{displaymath}
so that we see the $V_3$ locus is interpreted in this case as the
locus where the quantum ring relations become ill-defined (which is
also the discriminant locus).

Now, let us compute the locus $V_4$.  Proceeding as above, we find
\begin{displaymath}
\left[ \begin{array}{c} R_{(4)} \\ R_{(3,1)} \\ R_{(2,2)}
\end{array} \right] \: = \: (C^4_{\mu \nu}) \left[ \begin{array}{c}
\sigma_{(4)} \\ \sigma_{(3,1)} \\ \sigma_{(2,2)} \end{array}
\right],
\end{displaymath}
where
\begin{displaymath}
(C^4_{\mu \nu}) \: = \: \left[ \begin{array}{ccc}
1 + I_1 + I_2 + I_3 & I_1 + I_2 + I_3 & I_2 + 2 I_3 \\
0 & 1 & I_1 + I_2 \\
-I_3 & -1 -I_3 & 1 - I_2 - 2 I_3
\end{array} \right]
\end{displaymath}
and in particular we find
\begin{displaymath}
\det C^4_{\mu \nu} \: = \: \det C^3_{\mu \nu},
\end{displaymath}
so that the locus
\begin{displaymath}
V_4 \: = \: V_3 \: = \: \: \{ \Delta = 0 \}.
\end{displaymath}
As the discriminant locus already appears to exhaust the ways in
which the quantum cohomology ring can degenerate, this result should
not be surprising.

\subsection{$G(2,4)$}

\subsubsection{(2,2) theory}

Let's first compute the correlation functions for the (2,2) theory
engineering $G(2,4)$. By computing the correlation functions, we
want to explicitly show that
\[
R_{(3)} = \sigma_{\tiny\yng(3)}=0,~ \: \: R_{(4)} + q =
\sigma_{\tiny\yng(4)} + q = 0,
\]
as implied by our general result. The four-point correlation
functions in the theory are given by
\begin{displaymath}
\langle f(\sigma) \rangle \: = \: \frac{1}{2!} {\rm JKG-Res}\,
\left\{ - (\sigma_1-\sigma_2)^2 \frac{1}{\sigma_1^4}
\frac{1}{\sigma_2^4} f(\sigma) \right\}
\end{displaymath}
The Jeffrey-Kirwan-Grothendieck residues in this case are merely
iterated ordinary contour integrals about $\sigma_1=0$ and
$\sigma_2=0$, {\it i.e.}
\begin{displaymath}
\langle f(\sigma) \rangle \: = \: \frac{1}{2!} \oint d \sigma_2
\oint d \sigma_1  \left\{ - (\sigma_1-\sigma_2)^2
\frac{1}{\sigma_1^4} \frac{1}{\sigma_2^4} f(\sigma) \right\}
\end{displaymath}
It is straightforward to show that
\[
\begin{split}
\langle \sigma_1^4 \rangle = 0,~~&
\langle \sigma_1^3 \sigma_2 \rangle = - \frac{1}{2!}, \\
\langle \sigma_1^2 \sigma_2^2 \rangle = + \frac{2}{2!},~~&
\langle \sigma_1 \sigma_2^3 \rangle = - \frac{1}{2!}, \\
\langle \sigma_2^4 \rangle = 0.~~&
\end{split}
\]

Now, let us interpret this in terms of the cohomology of $G(2,4)$.
In principle, the cohomology classes correspond to Young tableaux
sitting inside the $2 \times (4-2)$ box
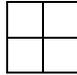
\begin{displaymath}
\yng(2,2)
\end{displaymath}
and so in particular are given by
\begin{eqnarray*}
\sigma_{\tiny\yng(1)} & = & \sigma_1 + \sigma_2 \\
\sigma_{\tiny\yng(2)} & = & \sigma_1^2 + \sigma_2^2 + \sigma_1 \sigma_2 \\
\sigma_{\tiny\yng(1,1)} & = & \sigma_1 \sigma_2 \\
\sigma_{\tiny\yng(2,1)} & = & \sigma_1^2 \sigma_2 + \sigma_1 \sigma_2^2 \\
\sigma_{\tiny\yng(2,2)} & = & \sigma_1^2 \sigma_2^2
\end{eqnarray*}
with relations, for example
\begin{displaymath}
\sigma_{\tiny\yng(3)} \: = \: \sigma_1^3 + \sigma_2^3 + \sigma_1^2
\sigma_2 + \sigma_1 \sigma_2^2 \: = \: 0
\end{displaymath}
Let us check that the relation $R_3 = \sigma_{\tiny\yng(3)}=0$ is
encoded in the correlation functions above.  Since it involves
third-order powers, and the correlation functions involve
fourth-order powers, we need to multiply by single copies of
$\sigma_1$, $\sigma_2$.  In other words, we claim the following
statements should be true:
\begin{displaymath}
\langle \sigma_1 \sigma_{\tiny\yng(3)} \rangle \: = \: 0 \: = \:
\langle \sigma_2 \sigma_{\tiny\yng(3)} \rangle
\end{displaymath}
Explicitly,
\begin{eqnarray*}
\langle \sigma_1 \sigma_{\tiny\yng(3)} \rangle & = & \langle
\sigma_1 ( \sigma_1^3 + \sigma_2^3 + \sigma_1^2 \sigma_2 +
\sigma_1 \sigma_2^2 ) \rangle \\
& = & \langle \sigma_1^4 \rangle + \langle \sigma_1 \sigma_2^3
\rangle + \langle \sigma_1^3 \sigma_2 \rangle + \langle \sigma_1^2
\sigma_2^2 \rangle
\end{eqnarray*}
and it is easy to check that this does indeed vanish. One can
similarly verify $\langle \sigma_2 \sigma_{\tiny\yng(3)} \rangle =
0$.

Correlation functions in the one-instanton sector are of the form
\begin{eqnarray*}
\langle f(\sigma) \rangle & = & \frac{1}{2!} {\rm JKG-Res}\left\{ q
(\sigma_1 - \sigma_2)^2 \frac{1}{\sigma_1^8} \frac{1}{\sigma_2^4}
f(\sigma) \right\} \\
& & + \frac{1}{2!} {\rm JKG-Res} \left\{ q (\sigma_1-\sigma_2)^2
\frac{1}{\sigma_1^4} \frac{1}{\sigma_2^8} f(\sigma) \right\}
\end{eqnarray*}
(where the JKG residues again reduce to iterated ordinary contour
integrals about $\sigma_1=0$ and $\sigma_2=0$), from which we
compute that the nonvanishing correlation functions are
\begin{displaymath}
\langle \sigma_1^5 \sigma_2^3 \rangle \: = \: q/2 \: = \: \langle
\sigma_1^7 \sigma_2 \rangle, \: \: \: \langle \sigma_1^6 \sigma_2^2
\rangle \: = \:  -q,
\end{displaymath}
\begin{displaymath}
\langle \sigma_1^3 \sigma_2^5 \rangle \: = \: q/2 \: = \: \langle
\sigma_1 \sigma_2^7 \rangle, \: \: \: \langle \sigma_1^2 \sigma_2^6
\rangle \: = \: -q.
\end{displaymath}

Using the fact that
\[
\sigma_{\tiny\yng(4)}  =  \sigma_1^4 \: + \: \sigma_1^3 \sigma_2 \:
+ \: \sigma_1^2 \sigma_2^2 \: + \: \sigma_1 \sigma_2^3 \: + \:
\sigma_2^4,
\]
we compute
\begin{displaymath}
\langle \sigma_{\tiny\yng(4)} \rangle \: = \: 0
\end{displaymath}
\begin{displaymath}
\langle \sigma_1^4 \sigma_{\tiny\yng(4)} \rangle \: = \: 0 \: = \:
\langle \sigma_2^4 \sigma_{\tiny\yng(4)} \rangle
\end{displaymath}
\begin{displaymath}
\langle \sigma_1^3 \sigma_2 \sigma_{\tiny\yng(4)} \rangle \: = \:
q/2 \: = \: \langle \sigma_1 \sigma_2^3 \sigma_{\tiny\yng(4)}
\rangle
\end{displaymath}
\begin{displaymath}
\langle \sigma_1^2 \sigma_2^2 \sigma_{\tiny\yng(4)} \rangle \: = \:
-q
\end{displaymath}
From the expressions
\begin{displaymath}
\sigma_{\tiny\yng(3,1)} = \sigma_1^3 \sigma_2 \: + \: \sigma_1^2
\sigma_2^2 \: + \: \sigma_1 \sigma_2^3
\end{displaymath}
\begin{displaymath}
\sigma_{\tiny\yng(2,2)} = \sigma_1^2 \sigma_2^2
\end{displaymath}
we get
\begin{displaymath}
\langle \sigma_{\tiny\yng(4)} ~\sigma_{\tiny\yng(4)} \rangle = 0
\end{displaymath}
\begin{displaymath}
\langle \sigma_{\tiny\yng(4)} ~\sigma_{\tiny\yng(3,1)} \rangle = 0
\end{displaymath}
\begin{displaymath}
\langle \sigma_{\tiny\yng(4)} ~\sigma_{\tiny\yng(2,2)} \rangle = -q
\end{displaymath}
Thus we see the relation $R_{(4)}$ is valid because
$\sigma_{\tiny\yng(4)} = \langle \sigma_{\tiny\yng(4)}
~\sigma_{\tiny\yng(2,2)} \rangle \cdot 1 = -q$. The other relation
at fourth order can be read off immediately,
\[
\sigma_{\tiny\yng(3,1)} =
\sigma_{\tiny\yng(3)}~\sigma_{\tiny\yng(1)} - \sigma_{\tiny\yng(4)}
= q.
\]

\subsubsection{(0,2) theory}

We can study the (0,2) theories following the same procedure. Again,
after absorbing a sign in $q$, the localization formula
\eqref{localization(0,2)} reads
\begin{eqnarray*}
\lefteqn{ \langle f(\sigma) \rangle \: =
}\\
& & \frac{1}{2!} \sum_{{\mathfrak m}_1,\cdots,{\mathfrak m}_k \in
{\mathbb Z} } {\rm JKG-Res} \, \left\{ (-q)^{\sum {\mathfrak m}_i}
\left( (-) (\sigma_{1} - \sigma_{2} ) \right) \prod_{a=1}^2
\left( \frac{1}{\det \tilde{E}(\sigma_{a})} \right)^{{\mathfrak
m}_i+1} f(\sigma) \right\},
\end{eqnarray*}
where
\begin{displaymath}
\tilde{E}^i_j(\sigma_{a}) \: = \: \delta^i_j \sigma_{a} \:
+ \: B^i_j \left( \sum_{b} \sigma_{b} \right).
\end{displaymath}

In this case, the JKG residue gives us the following iterated
residue prescription for generic $b_j$:
\begin{enumerate}
\item First, we perform a contour integral over $\sigma_1$, summing
over the residues at the four loci
\begin{displaymath}
\sigma_1 \: = \: - \sigma_2 \frac{b_j}{1 + b_j}
\end{displaymath}
for $j \in \{1,2,3,4\}$, corresponding to the roots of $\det
\tilde{E}(\sigma_1)$,
\item then, we perform a contour integral over $\sigma_2$, taking the
residue at $\sigma_2=0$.
\end{enumerate}
In this case, the results for the classical correlation functions
(${\mathfrak m}_1 = 0 = {\mathfrak m}_2$) are as follows:
\begin{eqnarray*}
\langle \sigma_1^4 \rangle & = & \Delta^{-1}\left[ I_1 + 2 I_1^2 + 4
I_1 I_2 - 2 I_3 + 2 I_2^2 + 2 I_1 I_3 - 4 I_4 + 2 I_2 I_3
- 2 I_1 I_4 \right], \\
\langle \sigma_1^3 \sigma_2 \rangle & = & \Delta^{-1} \left[ -1 -3
I_1 - 2 I_1^2 - 3 I_2 - 4 I_1 I_2 -2 I_2^2 - I_3 - 2 I_1 I_3 + 4 I_4
- 2 I_2 I_3 + 2 I_1 I_4
\right], \\
\langle \sigma_1^2 \sigma_2^2 \rangle & = & \Delta^{-1} \left[ 2 + 4
I_1 + 2 I_1^2 + 4 I_2 + 4 I_1 I_2 + 2 I_3 - 4 I_4 + 2 I_2^2 + 2 I_1
I_3 + 2 I_2 I_3 -2 I_1 I_4
\right], \\
\langle \sigma_1 \sigma_2^3 \rangle & = & \langle \sigma_1^3
\sigma_2 \rangle,
 \\
\langle \sigma_2^4 \rangle & = & \langle \sigma_1^4 \rangle,
\end{eqnarray*}
or
\begin{eqnarray*}
\langle \sigma_{\tiny\yng(4)} \rangle & = & \langle \sigma_1^4
\rangle \: + \: \langle \sigma_1^3 \sigma_2 \rangle \: + \: \langle
\sigma_1^2 \sigma_2^2 \rangle \: + \:
\langle \sigma_1 \sigma_2^3 \rangle \: + \: \langle \sigma_2^4 \rangle, \\
& = & 2 \Delta^{-1} \left[ - I_2 +  I_1^2 + 2 I_2 I_1 -2 I_3 + I_2^2
-2 I_4 + I_1 I_3 - I_1 I_4 +  I_2 I_3 \right],
\end{eqnarray*}
\begin{eqnarray*}
\langle \sigma_{\tiny\yng(2)}^2 \rangle & = & \langle \sigma_1^4
\rangle \: + \: 2 \langle \sigma_1^3 \sigma_2 \rangle \: + \: 3
\langle \sigma_1^2 \sigma_2^2 \rangle \: + \: 2 \langle \sigma_1
\sigma_2^3 \rangle \: + \:
\langle \sigma_2^4 \rangle, \\
& = & \Delta^{-1} \left[ 2 + 2 I_1 + 2 I_1^2 + 4 I_1 I_2 -2 I_3 - 4
I_4 + 2 I_2^2 + 2 I_1 I_3 + 2 I_2 I_3 - 2 I_1 I_4 \right],
\end{eqnarray*}
\begin{eqnarray*}
\langle \sigma_{\tiny\yng(1)} \sigma_{\tiny\yng(3)} \rangle & = &
\langle \sigma_1^4 \rangle \: + \: 2 \langle \sigma_1^3 \sigma_2
\rangle \: + \: 2 \langle \sigma_1^2 \sigma_2^2 \rangle \: + \: 2
\langle \sigma_1 \sigma_2^3 \rangle \: + \:
\langle \sigma_2^4 \rangle, \\
& = & \Delta^{-1} \left[ -2 I_1 - 4 I_2 - 4 I_3 \right] ,
\end{eqnarray*}
\begin{eqnarray*}
\langle \sigma_{\tiny\yng(1)}^2 \sigma_{\tiny\yng(2)} \rangle & = &
\langle \sigma_1^4 \rangle \: + \: 3 \langle \sigma_1^3 \sigma_2
\rangle \: + \: 4 \langle \sigma_1^2 \sigma_2^2 \rangle \: + \: 3
\langle \sigma_1 \sigma_2^3 \rangle \: + \:
\langle \sigma_2^4 \rangle, \\
& = & \Delta^{-1} \left[ 2 - 2 I_2 - 2 I_3 \right],
\end{eqnarray*}
\begin{eqnarray*}
\langle \sigma_{\tiny\yng(1)}^4 \rangle & = & \langle \sigma_1^4
\rangle \: + \: 4 \langle \sigma_1^3 \sigma_2 \rangle \: + \: 6
\langle \sigma_1^2 \sigma_2^2 \rangle \: + \: 4 \langle \sigma_1
\sigma_2^3 \rangle \: + \:
\langle \sigma_2^4 \rangle, \\
& = & \Delta^{-1} \left[ 4 + 2 I_1 \right],
\end{eqnarray*}
\begin{eqnarray*}
\langle \sigma_{\tiny\yng(3,1)} \rangle & = & \langle \sigma_1^3
\sigma_2 \rangle \: + \: \langle \sigma_1^2 \sigma_2^2 \rangle \: +
\:
\langle \sigma_1 \sigma_2^3 \rangle, \\
& = & \Delta^{-1} \left[ -2 I_1 -2 I_2 -2 I_1^2 + 4 I_4 -2 I_2^2 -2
I_3 I_1 -2 I_3 I_2 + 2 I_1 I_4 -4 I_2 I_1 \right],
\end{eqnarray*}
\begin{eqnarray*}
\langle \sigma_{\tiny\yng(2,2)} \rangle & = &
\langle \sigma_1^2 \sigma_2^2 \rangle, \\
& = & \Delta^{-1} \left[ 2 + 4 I_1 + 2 I_1^2 + 4 I_2 + 4 I_1 I_2 + 2
I_3 - 4 I_4 + 2 I_2^2 + 2 I_1 I_3 + 2 I_2 I_3 -2 I_1 I_4 \right],
\end{eqnarray*}
where the characteristic polynomials of $B$ are given explicitly as
\begin{eqnarray*}
I_1 & = & \sum_i b_i \: = \: {\rm tr}\, B, \\
I_2 & = & \sum_{i < j} b_i b_j, \\
I_3 & = & \sum_{i < j < k} b_i b_j b_k, \\
I_4 & = & b_1 b_2 b_3 b_4 \: = \: \det B,
\end{eqnarray*}
and
\begin{eqnarray*}
\Delta  & = &
2 \prod_{i < j} \left( 1 + b_i + b_j \right), \\
& = & 2 \left( 1 + 3 I_1 + 3 I_1^2 + 2 I_2 + I_1^3 + 4 I_1 I_2 + 2
I_1^2 I_2 + I_2^2 + I_1 I_3 - 4 I_4 + I_1 I_2^2
\right. \\
& & \hspace*{1.5in} \left.
 + I_1^2 I_3 +
I_1 I_2 I_3 - 4 I_1 I_4 - I_1^2 I_4 - I_3^2 \right) .
\end{eqnarray*}
We see the discriminant locus is given by $\Delta = 0$, this is
consistent with our general result in \cite{3}, which says that the
$B$-deformation fails to define a vector bundle on $G(k,n)$ if and
only if there exits $k$ eigenvalues of $B$ whose sum is $-1$.

Now, the quantum sheaf cohomology ring for this model is predicted
by~(\ref{eq:general-qsc-ring}) to be
\begin{displaymath}
{\mathbb C}\left[\sigma_{(1)}, \sigma_{(2)}, \cdots\right] /
\left\langle D_3, D_4, \cdots, R_{(3)}, R_{(4)}+q, R_{(5)} + q
\sigma_{(1)}, \cdots
 \right\rangle,
\end{displaymath}
where
\begin{eqnarray*}
R_{(3)} & = & \sum_{i=0}^3 I_i \sigma_{(3-i)} \sigma_{(1)}^i, \\
& = & \sigma_{(3)} + I_1 \sigma_{(2)} \sigma_{(1)} +
(I_2 + I_3) \sigma_{(1)}^3, \\
R_{(4)} & = & \sum_{i=0}^4 I_i \sigma_{(r-i)} \sigma_{(1)}^i , \\
& = & \sigma_{(4)} + I_1 \sigma_{(3)} \sigma_{(1)} + I_2
\sigma_{(2)} \sigma_{(1)}^2 + (I_3 + I_4) \sigma_{(1)}^4.
\end{eqnarray*}

As a consistency test, it is straightforward to check that the
relations above are reflected in the correlation functions.  For
example, the classical correlation functions are easily demonstrated
to obey
\begin{displaymath}
\langle \sigma_{\tiny\yng(1)} D_3 \rangle \: = \: \langle D_4
\rangle \: = \: \langle \sigma_{\tiny\yng(1)} R_{(3)} \rangle \: =
\: 0.
\end{displaymath}
Now, we also have the relation $R_{(4)}=-q$, which for the purely
classical correlation functions implies
\begin{displaymath}
\langle R_{(4)} \rangle \: = \: 0,
\end{displaymath}
which is also easily checked to be true.  By including instanton
sectors, one can see the full quantum-corrected relation, as we
shall discuss next.


The relation $R_{(4)}=-q$ can be derived from the quantum cohomology
ring relation derived from the Jeffrey-Kirwan-Grothendieck residue
expression, namely,
\begin{displaymath}
\det \tilde{E}(\sigma_1) \: = \: -q \: = \: \det \tilde{E}(\sigma_2)
,
\end{displaymath}
where
\begin{displaymath}
\tilde{E}(x) \: = \: I x \: + \:  B (\sigma_1 + \sigma_2) .
\end{displaymath}

Now, it is straightforward to expand
\begin{displaymath}
\det \tilde{E}(x) \: = \: x^4 \: + \: I_1 (\sigma_1 + \sigma_2) x^3
\: + \: I_2 (\sigma_1 + \sigma_2)^2 x^2 \: + \: I_3 (\sigma_1 +
\sigma_2)^3 x \: + \: I_4 (\sigma_1 + \sigma_2)^4
\end{displaymath}
so
\begin{eqnarray*}
\langle \det \tilde{E}(\sigma_1) \rangle & = & \langle \sigma_1^4
\rangle ( 1 + I_1 + I_2 + I_3 + 2 I_4) \: + \: \langle \sigma_1^3
\sigma_2 \rangle ( I_1 + 2 I_2 + 4 I_3 + 8 I_4)
\\
& & \: + \: \langle \sigma_1^2 \sigma_2^2 \rangle (I_2 + 3 I_3 + 6
I_4) ,
\\
& = & \langle \det \tilde{E}(\sigma_2) \rangle ,
\end{eqnarray*}
which implies
\[
\begin{split}
2 R_{(4)} - R_3 \sigma_{\tiny\yng(1)} \: = \:& \sigma_{\tiny\yng(4)}
(1 + I_1 + I_2 + I_3 + 2 I_4) \: + \: \langle
\sigma_{\tiny\yng(3,1)} \rangle (-1 + I_2 + 3 I_3 + 6 I_4)
 \\
&\: + \: \langle \sigma_{\tiny\yng(2,2)} \rangle (-I_1 + 2 I_3 + 4
I_4) \: = \: -2q,
\end{split}
\]
or simply $R_{(4)} + q = 0$ as expected.

Now, let us turn to the interpretation of the loci $V_m$ in this
example. First, consider the $V_3$ locus.  It is straightforward to
compute
\begin{displaymath}
\left[ R_{(3)}\right] \: = \: \left[ 1+ I_1 + I_2 + I_3, I_1 + 2I_2
+ 2I_3 \right]\left[ \begin{array}{c} \sigma_{(3)} \\ \sigma_{(2,1)}
\end{array} \right],
\end{displaymath}
hence
\begin{eqnarray*}
V_3 & = & \left\{ 1+I_1+I_2+I_3 = 0 \mbox{ and }
I_1 + 2I_2 + 2I_3 = 0 \right\}, \\
& = & \{ I_2+I_3 = +1, \: I_1 = -2 \}.
\end{eqnarray*}
Note that this locus does not intersect the (2,2) locus, as expected
on general grounds, as the $I_i$ never all become zero.

Next, we compute the $V_4$ locus. It is straightforward to compute
\begin{displaymath}
\left[ \begin{array}{c} R_{(4)} \\ R_{(3,1)} \end{array} \right] \:
= \: (C^4_{\mu \nu}) \left[ \begin{array}{c} \sigma_{(4)} \\
\sigma_{(3,1)} \\ \sigma_{(2,2)} \end{array} \right],
\end{displaymath}
where
\begin{displaymath}
(C^4_{\mu \nu}) \: = \: \left[
\begin{array}{ccc}
1 + I_1 + I_2 + I_3 + I_4 & I_1 + 2 I_2 + 3 I_3 + 3 I_4 & I_2 + 2 I_3 + 2 I_4 \\
-I_4 & 1 + I_1 + I_2 - 3 I_4 & I_1 + I_2 - 2 I_4
\end{array} \right] .
\end{displaymath}
Let $M_{12}$, $M_{13}$, $M_{23}$ denote the three $2\times 2$
submatrices of $(C^4_{\mu \nu})$ formed by omitting a column, then
the locus $V_4$ is defined as
\begin{displaymath}
V_4 \: = \: \{ M_{12} = 0 = M_{13} = M_{23} \}.
\end{displaymath}

For completeness, we also list here results for $V_5$ and $V_6$.
First, $V_5$ is computed from the relation
\[ \left[
\begin{array}{c}
R_{(5)} \\ R_{(4,1)} \\ R_{(3,2)}
\end{array} \right] =
\left( C^5_{\mu \nu}
 \right)
\left[
\begin{array}{c}
\sigma_{(5)} \\ \sigma_{(4,1)} \\ \sigma_{(3,2)}
\end{array} \right]
\]
for
\begin{displaymath}
(C^5_{\mu \nu}) \: = \: \left[
\begin{array}{ccc}
1 + I_1 + I_2 + I_3 + I_4 & I_1 + 2 I_2 + 3 I_3 + 4 I_4 & I_2 + 3 I_3 + 5 I_4  \\
0 & 1 + I_1 + I_2 + I_3 & I_1 + 2 I_2 + 2 I_3 \\
-I_4 & -I_3 - 4 I_4 & 1 + I_1 - 2 I_3 - 5 I_4
\end{array} \right],
\end{displaymath}
and $V_6$ is computed from the relation
\[ \left[
\begin{array}{c}
R_{(6)} \\ R_{(5,1)} \\ R_{(4,2)} \\ R_{(3,3)}
\end{array} \right] =
\left( C^6_{\mu \nu} \right) \left[
\begin{array}{c}
\sigma_{(6)} \\ \sigma_{(5,1)} \\ \sigma_{(4,2)} \\ \sigma_{(3,3)}
\end{array} \right]
\]
for $(C^6_{\mu \nu})$ given by
\begin{displaymath}
 \left[
\begin{array}{cccc}
1 + I_1 + I_2 + I_3 + I_4 & I_1 + 2 I_2 + 3 I_3 + 4 I_4 & I_2 + 3 I_3 + 6 I_4 & I_3 + 3 I_4 \\
0 & 1 + I_1 + I_2 + I_3 + I_4 & I_1 + 2 I_2 + 3 I_3 + 3 I_4 & I_2 + 2 I_3 + 2 I_4 \\
0 & -I_4 & 1 + I_1 + I_2 - 3 I_4 & I_1 + I_2 - 2 I_4 \\
-I_4 & -I_3 - 4 I_4 & -I_2 - 3 I_3 - 6 I_4 & 1 - I_2 - 2 I_3 - 3 I_4
\end{array} \right].
\end{displaymath}
At least when $B$ is diagonal, it is straightforward to check that
\begin{displaymath}
\det(C_5) \: = \: \det(C_6) \: = \: \Delta,
\end{displaymath}
or equivalently,
\begin{displaymath}
V_5 \: = \: V_6 \: = \: X,
\end{displaymath}
consistent with our expectation that for $m$ larger than the
dimension of the Grassmannian, $V_m$ matches the discriminant locus.

\subsection{$G(2,5)$}

In the previous sections we described the results for the
Grassmannian $G(2,4)$.  Although this is not a projective space, it
can be described as a hypersurface in a projective space, so in this
section we give one additional nonabelian example, one which is not related or
dual to an abelian GLSM, to demonstrate
the results.  Specifically, in this section we will consider the theory
for $G(2,5)$. 

\subsubsection{(2,2) theory}

The (classical) six-point correlation functions are given by
\begin{displaymath}
\langle f(\sigma) \rangle \: = \: \frac{1}{2!} {\rm JKG-Res} \left\{
(-) (\sigma_1 - \sigma_2)^2 \frac{1}{\sigma_1^5}
\frac{1}{\sigma_2^5} f(\sigma) \right\}.
\end{displaymath}
We can compute these as iterated ordinary contour integrals about
$\sigma_1 = 0$ and $\sigma_2 = 0$.  It is straightforward to show
that
\[
\begin{split}
\langle \sigma_1^6 \rangle  = 0,~~& \langle \sigma_1^5 \sigma_2
\rangle  =
0, \\
\langle \sigma_1^4 \sigma_2^2 \rangle = - \frac{1}{2!},~~& \langle
\sigma_1^3 \sigma_2^3 \rangle =
+ \frac{2}{2!}, \\
\langle \sigma_1^2 \sigma_2^4 \rangle = - \frac{1}{2!},~~& \langle
\sigma_1 \sigma_2^5 \rangle =
0,\\
\langle \sigma_2^6 \rangle = 0.~~&
\end{split}
\]

All the nonzero cohomology classes should be defined by Young
diagrams fitting inside the $2 \times 3$ box
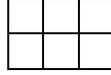
\begin{displaymath}
\yng(3,3)
\end{displaymath}
Using the correlation functions above, it is straightforward to
compute the ring relations
\begin{eqnarray*}
\sigma_{\tiny\yng(4)} & = & \sigma_1^4 + \sigma_1^3 \sigma_2 +
\sigma_1^2 \sigma_2^2 +
\sigma_1 \sigma_2^3 + \sigma_2^4 \: = \: 0 \\
\sigma_{\tiny\yng(4,1)} & = & \sigma_1^4 \sigma_2 + \sigma_1^3
\sigma_2^2 + \sigma_1^2 \sigma_2^3 +
\sigma_1 \sigma_2^4 \: = \: 0 \\
\sigma_{\tiny\yng(4,2)} & = & \sigma_1^4 \sigma_2^2 + \sigma_1^3
\sigma_2^3 + \sigma_1^2 \sigma_2^4 \: = \: 0
\end{eqnarray*}
which matches the ring relations one expects from the cohomology
theory. In each case, one multiplies in arbitrary powers of
$\sigma_1$, $\sigma_2$ to get a six-point function, and in each
case, the sum amounts to a scan through values that sum to zero.
The top-form, described by $\sigma_{\tiny\yng(3,3)} = \sigma_1^3
\sigma_2^3$, has nonzero vev, as expected.

Correlation functions in the one-instanton sector are of the form
\begin{eqnarray*}
\lefteqn{ \langle f(\sigma) \rangle  \: =
} \\
& &  \frac{1}{2!} {\rm JKG-Res}\left\{ q (\sigma_1 - \sigma_2)^2
\frac{1}{\sigma_1^{10}} \frac{1}{\sigma_2^5} f(\sigma) \right\} +
\frac{1}{2!} {\rm JKG-Res} \left\{ q (\sigma_1-\sigma_2)^2
\frac{1}{\sigma_1^5} \frac{1}{\sigma_2^{10}} f(\sigma) \right\} ,
\end{eqnarray*}
from which one can compute the nonzero correlation functions at
order 11 to be
\[
\begin{split}
\langle \sigma_1^9 \sigma_2^2 \rangle = \frac{q}{2}, \: \: \:
\langle \sigma_1^8 \sigma_2^3 \rangle = -q, \: \: \:
\langle \sigma_1^7 \sigma_2^4 \rangle = \frac{q}{2}, \\
\langle \sigma_1^4 \sigma_2^7 \rangle = \frac{q}{2}, \: \: \:
\langle \sigma_1^3 \sigma_2^8 \rangle = -q, \: \: \: \langle
\sigma_1^2 \sigma_2^9 \rangle = \frac{q}{2}.
\end{split}
\]
Thus we see
\[
\langle \sigma_{\tiny\yng(5)}~ \sigma_{\tiny\yng(3,3)} \rangle =
-q,~ \: \: \langle \sigma_{\tiny\yng(6)}~ \sigma_{\tiny\yng(3,2)}
\rangle = -q,
\]
which impies
\[
R_{(5)} + q = 0,~ \: \: R_{(6)} + q \sigma_{\tiny\yng(1)} = 0
\]
because
\[
\langle \sigma_{\tiny\yng(3,3)} \rangle = 1, \: \: \: \langle
\sigma_{\tiny\yng(3,2)}~ \sigma_{\tiny\yng(1)} \rangle = 1.
\]

\subsubsection{(0,2) theory}

Next, let us consider the (0,2) theory defined by deformations of
the tangent bundle.

In this case, classical correlation functions are given by
\begin{eqnarray*}
\lefteqn{ \langle f(\sigma) \rangle \: = \:
} \\
& & \frac{1}{2!} {\rm JKG-Res}\left\{ (-) (\sigma_1 - \sigma_2)^2
\left( \frac{1}{\det \tilde{E}(\sigma_1) } \right) \left(
\frac{1}{\det \tilde{E}(\sigma_2) } \right) f(\sigma) \right\},
\end{eqnarray*}
and for $B$ diagonal, computing much as in previous examples, we
find:
\begin{eqnarray*}
\langle \sigma_1^6 \rangle & = & \Delta^{-1} \left[ - I_1^2 + I_2
- 2 I_1^3 + I_1 I_2 + I_3 + 3 I_2^2 + 4 I_1 I_3 - I_4 - 6 I_1^2 I_2
- 6 I_1 I_2^2 - 4 I_1^2 I_3
\right. \\
& & \: \: \: \: \: \left.
 + 10 I_2 I_3 - 8 I_1 I_2 I_3
-2 I_2^3 + 7 I_3^2 -4 I_2^2 I_3 + 5 I_4 I_1 - 5 I_5 + 4 I_1 I_5
\right. \\
& & \: \: \: \: \: \left.
 + 8 I_1^2 I_5
+ 9 I_4 I_2 - 2 I_1 I_3^2 - 2 I_2 I_3^2 - 2 I_1 I_2 I_4 - 2 I_2^2
I_4 + 11 I_3 I_4 - 2 I_2 I_3 I_4
\right. \\
& & \: \: \: \: \: \left.
 + 4 I_4^2 + 2 I_1 I_4^2 + I_2 I_5
+ 8 I_1 I_2 I_5 + 2 I_2^2 I_5 - 2 I_3 I_5 - 2 I_4 I_5 \right],
\end{eqnarray*}
\begin{eqnarray*}
\langle \sigma_1^5 \sigma_2 \rangle & = & \Delta^{-1} \left[ I_1 + 3
I_1^2 + 2 I_1^3 + 5 I_1 I_2 - 2 I_3 + 6 I_1^2 I_2 + 2 I_2^2 + I_1
I_3 - 4 I_4
+ 6 I_1 I_2^2
\right. \\
& & \: \: \: \: \: \left.
 + 4 I_1^2 I_3 - 5 I_1 I_4 -5 I_5
+ 2 I_2^3 - 2 I_3^2 + 8 I_1 I_2 I_3 -4 I_2 I_4 -14 I_1  I_5
\right. \\
& & \: \: \: \: \: \left.
+ 4 I_2^2 I_3 + 2 I_1 I_3^2 + 2 I_1 I_2 I_4 - 6 I_3 I_4 - 8 I_1^2
I_5 - 6 I_2 I_5
+ 2 I_2 I_3^2 + 2 I_2^2 I_4
\right. \\
& & \: \: \: \: \: \left.
 - 4 I_4^2 - 8 I_1 I_2 I_5 + 2 I_3 I_5
+ 2 I_2 I_3 I_4 -2 I_1 I_4^2 -2 I_2^2 I_5 + 2 I_4 I_5 \right],
\end{eqnarray*}
\begin{eqnarray*}
\langle \sigma_1^4 \sigma_2^2 \rangle & = & \Delta^{-1} \left[
-1 -4 I_1 - 5 I_1^2 -4 I_2 - 2 I_1^3 - 10 I_1 I_2 -2 I_3
-6 I_1^2 I_2 - 5 I_2^2 -6 I_1 I_3 + 3 I_4
\right. \\
& & \: \: \: \left.
-6 I_1 I_2^2 -4 I_1^2 I_3
 -6 I_2 I_3 + 3 I_1 I_4 + 13 I_5
-2 I_2^3 -8 I_1 I_2 I_3 - I_3^2 + I_2 I_4
\right. \\
& & \: \: \: \: \: \left.
 + 20 I_1 I_5
-4 I_2^2 I_3 -2 I_1 I_3^2 -2 I_1 I_2 I_4 + 3 I_3 I_4 + 8 I_1^2 I_5 +
9 I_2 I_5
-2 I_2 I_3^2
\right. \\
& & \: \: \: \: \: \left.
 -2 I_2^2 I_4 + 4 I_4^2 + 8 I_1 I_2 I_5 - 2 I_3 I_5
- 2 I_2 I_3 I_4 + 2 I_1 I_4^2 + 2 I_2^2 I_5 - 2 I_4 I_5 \right] ,
\end{eqnarray*}
\begin{eqnarray*}
\langle \sigma_1^3 \sigma_2^3 \rangle & = & \Delta^{-1}\left[
2 + 6 I_1 + 6 I_1^2 + 6 I_2 + 2 I_1^3 + 12 I_1 I_2 + 4 I_3
+ 6 I_1^2 I_2 + 6 I_2^2 - 2 I_4 + 8 I_1 I_3
\right. \\
& & \: \: \: \left.
+ 6 I_1 I_2^2 - 16 I_5 + 4 I_1^2 I_3 + 8 I_2 I_3 - 2 I_1 I_4
+ 2 I_2^3 - 22 I_1 I_5 + 8 I_1 I_2 I_3
\right. \\
& & \: \: \: \: \: \left. + 2 I_3^2
+ 4 I_2^2 I_3 + 2 I_1 I_3^2 + 2 I_1 I_2 I_3 - 2 I_3 I_4 - 2 I_1^2
I_5 - 10 I_2 I_5
+ 2 I_2 I_3^2
\right. \\
& & \: \: \: \: \: \left.
 + 2 I_2^2 I_4  - 4 I_4^2 - 8 I_1 I_2 I_5 + 2 I_3 I_5
+ 2 I_2 I_3 I_4 - 2 I_1 I_4^2 - 2 I_2^2 I_5  + 2 I_4 I_5 \right],
\end{eqnarray*}
\begin{displaymath}
\langle \sigma_1^2 \sigma_2^4 \rangle \: = \: \langle \sigma_1^4
\sigma_2^2 \rangle, \: \: \: \langle \sigma_1 \sigma_2^5 \rangle \:
= \: \langle \sigma_1^5 \sigma_2 \rangle, \: \: \: \langle
\sigma_2^6 \rangle \: = \: \langle \sigma_1^6 \rangle,
\end{displaymath}
where
\begin{eqnarray*}
\Delta & = & 2 \prod_{i < j} (1 + b_{i} + b_{j}), \\
\end{eqnarray*}
and
\begin{eqnarray*}
I_5 & = & b_1 b_2 b_3 b_4 b_5 \: = \: \det B, \\
I_4 & = & \sum_{i < j < k < \ell} b_{i} b_{j} b_{k} b_{\ell}, \\
I_3 & = & \sum_{i < j < k} b_{i} b_{j} b_{k}, \\
I_2 & = & \sum_{i < j} b_{i} b_{j}, \\
I_1 & = & \sum_i b_{i} \: = \: {\rm tr}\, B.
\end{eqnarray*}

The quantum sheaf cohomology ring~(\ref{eq:general-qsc-ring}) in
this case is given by
\begin{displaymath}
{\mathbb C}\left[ \sigma_{(1)}, \sigma_{(2)},\cdots \right] /
\left\langle D_3, D_4, \cdots,  R_{(4)}, R_{(5)} + q, R_{(6)} + q
\sigma_{(1)}, \cdots \right\rangle,
\end{displaymath}
where
\begin{eqnarray*}
R_{(4)} & = & \sum_{i=0}^{4} I_i \sigma_{(4-i)} \sigma_{(1)}^i, \\
& = & \sigma_{(4)} + I_1 \sigma_{(3)} \sigma_{(1)} +
I_2 \sigma_{(2)} \sigma_{(1)}^2 \: + \: (I_3 + I_4) \sigma_{(1)}^4, \\
R_{(5)} & = & \sum_{i=0}^5 I_i \sigma_{(5-i)} \sigma_{(1)}^i, \\
& = & \sigma_{(5)} + I_1 \sigma_{(4)} \sigma_{(1)} + I_2
\sigma_{(3)} \sigma_{(1)}^2 + I_3 \sigma_{(2)} \sigma_{(1)}^3
+ (I_4 + I_5) \sigma_{(1)}^5, \\
R_{(6)} & = & \sum_{i=0}^5 I_i \sigma_{(6-i)} \sigma_{(1)}^i, \\
& = & \sigma_{(6)} + I_1 \sigma_{(5)} \sigma_{(1)} + I_2
\sigma_{(4)} \sigma_{(1)}^2 + I_3 \sigma_{(3)} \sigma_{(1)}^3 + I_4
\sigma_{(2)} \sigma_{(1)}^4 + I_5 \sigma_{(1)}^6.
\end{eqnarray*}

As a consistency test, it is straightforward to check that the
relations above are reflected in the correlation functions.  For
example, the classical correlation functions are easily demonstrated
to obey
\begin{displaymath}
\langle \sigma_{\tiny\yng(1)}^3 D_3 \rangle \: = \: \langle
\sigma_{\tiny\yng(1)}^2 D_4 \rangle \: = \: \langle
\sigma_{\tiny\yng(1)} D_5 \rangle \: = \: \langle D_6 \rangle \: =
\: \langle \sigma_{\tiny\yng(1)}^2 R_{(4)} \rangle \: = \: 0.
\end{displaymath}
(Other vanishings of classical correlation functions are also
implied by the ring relations; for example, in the line above, one
could replace any instance of $\sigma_{\tiny\yng(1)}^2$ with
$\sigma_{\tiny\yng(2)}$ to get another vanishing correlation
function. Our intent above is merely to list some examples, not to
list every possible example.)

In addition, we can also use the classical correlation functions to
check the classical limits of the relations
\begin{displaymath}
R_{(5)} \: = \: -q, \: \: \: R_{(6)} \: = \: -q \sigma_{(1)}.
\end{displaymath}
In particular, it is straightforward to show that the classical
correlation functions obey
\begin{displaymath}
\langle \sigma_{\tiny\yng(1)} R_{(5)} \rangle \: = \: 0 \: = \:
\langle R_{(6)} \rangle,
\end{displaymath}
verifying the classical limit of the relations above.

One of the relations should be the classical limit of the quantum
cohomology ring relation derived from the
Jeffrey-Kirwan-Grothendieck residue expression, namely,
\begin{displaymath}
\det \tilde{E}(\sigma_1) \: = \: q \: = \: \det \tilde{E}(\sigma_2)
\end{displaymath}
where
\begin{displaymath}
\tilde{E}(x) \: = \: I x \: + \: B (\sigma_1 + \sigma_2)
\end{displaymath}
As before, it is straightforward to expand
\begin{displaymath}
\det \tilde{E}(x) \: = \: x^5 \: + \: I_1(\sigma_1 + \sigma_2) x^4
\: + \: I_2(\sigma_1 + \sigma_2)^2 x^3 \: + \: I_3(\sigma_1 +
\sigma_2)^3 x^2 \: + \: I_4(\sigma_1 + \sigma_2)^4 x \: + \:
I_5(\sigma_1 + \sigma_2)^5
\end{displaymath}
so, for example,
\begin{eqnarray*}
 \det \tilde{E}(\sigma_1)  & = &
 \sigma_1^5 (1 + I_1 + I_2 + I_3 + I_4 + I_5)
\: + \: \sigma_1^4 \sigma_2 ( I_1 + 2 I_2 + 3 I_3 + 4 I_4 + 5 I_5)
\\
& &  \: + \: \sigma_1^3 \sigma_2^2 ( I_2 + 3 I_3 + 6 I_4 + 10 I_5)
\: + \: \sigma_1^2 \sigma_2^3 ( I_3 + 4 I_4 + 10 I_5)
\\
& & \: + \: \sigma_1 \sigma_2^4 ( I_4 + 5 I_5) \: + \: \sigma_2^5 (
I_5)
\end{eqnarray*}
From this we derive
\begin{eqnarray*}
\lefteqn{ \sigma_{\tiny\yng(5)} \left( 1 + I_1 + I_2 + I_3 + I_4 + 2
I_5 \right) \: + \: \sigma_{\tiny\yng(4,1)} \left( -1 + I_2 + 2 I_3
+ 4 I_4 + 8 I_5 \right)
} \\
& & \hspace*{1.75in} \: + \: \sigma_{\tiny\yng(3,2)} \left( - I_1 -
I_2 + I_3 + 5 I_4 + 10 I_5 \right) \: = \: 2q
\end{eqnarray*}
It is straightforward to check, via multiplication by
$\sigma_{1,2}$, that the classical limit of the relation above is
indeed a property of the correlation functions given in this
section.

\section{Conclusions}

In this paper we have presented a proposal for the quantum sheaf
cohomology ring of Grassmannians with deformations of the tangent
bundle. We derived this proposal from one-loop effective actions,
and checked it in examples in which correlation functions were
computed with supersymmetric localization.  We also discussed where
this proposal is valid:  on codimension one subvarieties of the
space of tangent bundle deformations, not intersecting the (2,2)
locus, our proposal breaks down.  We discussed those loci
explicitly, and computed them in examples.

There are a number of questions that remain to be addressed. One
question concerns the role of duality.  In an ordinary Grassmannian,
$G(k,n)$ is the same space as $G(n-k,n)$, and both ordinary and
quantum cohomology of Grassmannians have presentations which respect
that symmetry.  By contrast, for tangent bundle deformations, our
presentation is not yet symmetric.  For example, the ring relations
for $G(1,n)$ take a very different form from those of $G(n-1,n)$.
Strictly speaking, the $B$ deformations encoded in
\begin{displaymath}
0 \: \longrightarrow \: S^* \otimes S \: \longrightarrow \: S^*
\otimes V \: \longrightarrow \: {\cal E} \: \longrightarrow \: 0,
\end{displaymath}
for ${\cal E}$ a deformation of the tangent bundle on $G(k,n)$,
dualize to deformations encoded in
\begin{displaymath}
0 \: \longrightarrow \: Q^* \otimes Q \: \longrightarrow \: Q
\otimes V^* \: \longrightarrow \: {\cal E}' \: \longrightarrow \: 0
\end{displaymath}
on $G(n-k,n)$, but the second sequence above does not have a simple
physical realization, and it is not immediately obvious how to
translate this into a parameter map.

Another open matter concerns the loci $V_m$.  We conjecture that on
$G(k,n)$, for $m > k(n-k)$, the $V_m$ are all identical to one
another and to the discriminant locus, so that there are only
finitely many components of the locus on which the quantum sheaf
cohomology ring relations break down, but we do not yet have a
proof.

Another open matter involves mathematical derivations.  A purely
mathematical derivation of the classical sheaf cohomology ring will
appear in \cite{3}.  An analogous mathematical derivation of the
quantum sheaf cohomology ring would technically involve sheaf theory
manipulations on Quot schemes, and is left for the future.

\section{Acknowledgements}

We would like to thank A.~Bertram, C.~Closset, R.~Donagi, W.~Gu,
B.~Jia, S.~Katz, I.~Melnikov, L.~Mihalcea, and D.~Park for useful
discussions.  This project began several years ago as joint work
with R.~Donagi and S.~Katz, and we thank them for their input.  Z.L.
was partially supported by EPSRC grant EP/J010790/1. E.S. was
partially supported by NSF grant PHY-1417410.

\appendix

\section{Mathematical representation}
\label{app:math-rep}

In this appendix we will summarize some of the ideas that will
appear in \cite{3}, including how representatives of sheaf
cohomology of deformations of the tangent bundle of $G(k,n)$ can be
represented by Young diagrams.

First, in order to determine the ring relations at order $r$, we
take the $r$th exterior power of \eqref{deformed2},
\[
\begin{split}
0 &\to \wedge^r \mathcal{E}^* \to \wedge^r(\mathcal{V}^* \otimes
\mathcal{S}) \to \wedge^{r-1}(\mathcal{V}^* \otimes \mathcal{S})
\otimes (\mathcal{S}^* \otimes \mathcal{S}) \to
\wedge^{r-2}(\mathcal{V}^* \otimes \mathcal{S}) \otimes
\mathrm{Sym}^2(\mathcal{S}^* \otimes \mathcal{S}) \\ &\to \cdots \to
\mathcal{V}^* \otimes \mathcal{S} \otimes
\mathrm{Sym}^{r-1}(\mathcal{S}^* \otimes \mathcal{S}) \to
\mathrm{Sym}^r(\mathcal{S}^* \otimes \mathcal{S}) \to 0.
\end{split}
\]
Breaking up the exact sequence above into $r$ short exact sequences,
we get a connecting map
\begin{displaymath}
\delta_r: \: H^0(\mathrm{Sym}^r(\mathcal{S}^* \otimes \mathcal{S}))
\: \longrightarrow \: H^r(\wedge^r \mathcal{E}^*)
\end{displaymath}
by composing the connecting maps associated with all the short exact
sequences. Thus the ring relations are encoded in the kernel of
$\delta_r$.\par

Now, $\mathrm{Sym}^r(\mathcal{S}^* \otimes \mathcal{S})$ can be
written as a direct sum in the form
$$\underset{\mu}{\oplus} \left(K_{\mu}
\mathcal{S^*} \otimes K_{\mu}\mathcal{S} \right),$$ where each $\mu$
is some Young diagram standing for an irreducible representation of
$U(k)$ and $K_{\mu}$ is the corresponding Schur functor. The direct
sum ranges over all the Young diagrams with $r$ boxes and at most
$k$ rows. Since
$$K_{\mu} \mathcal{S^*} \otimes K_{\mu}\mathcal{S} \: \cong \:
{\rm Hom}(K_{\mu}\mathcal{S}, K_{\mu}\mathcal{S}),$$ and each
$H^0(K_{\mu} \mathcal{S^*} \otimes K_{\mu}\mathcal{S})$ is
one-dimensional, a basis for $H^0(\mathrm{Sym}^r(\mathcal{S}^*
\otimes \mathcal{S}))$ can be taken to be
$$\{\sigma_\mu \mid \mu \mbox{ is a Young diagram with $r$ boxes and
at most $k$ rows } \},$$ where $\sigma_\mu$ is the identity bundle
map on $K_{\mu}\mathcal{S}$. \par

The product on
$$\underset{r\geqslant 0}{\oplus}
H^0 \left(\mathrm{Sym}^r(\mathcal{S}^* \otimes \mathcal{S})\right)$$
is defined to be the tensor product of bundle maps. Because
\[K_{\lambda} \mathcal{S} \otimes K_{\mu}\mathcal{S}
= \underset{\nu}{\bigoplus} N_{\lambda \mu \nu} K_{\nu}\mathcal{S},
\]
we see
\begin{equation}\label{product}
\sigma_{\lambda} \otimes \sigma_{\mu} = \underset{\nu}{\bigoplus}
N_{\lambda \mu \nu} \sigma_{\nu}.
\end{equation}
The numbers $N_{\lambda \mu \nu}$ are determined by the
Littlewood-Richardson rule. $N_{\lambda \mu \nu}$ is the number of
ways the Young diagram $\lambda$ can be expanded to the Young
diagram $\nu$ by a strict $\mu$-expansion. Note that \eqref{product}
would remain unchanged if one replaced each bundle map
$\sigma_{\mu}$ with the Schur polynomial corresponding to $\mu$, and
the tensor product with the usual product of polynomials. This
implies that the sheaf cohomology
$$\underset{r\geqslant 0}{\oplus}
H^0 (\mathrm{Sym}^r(\mathcal{S}^* \otimes \mathcal{S}))$$ is
isomorphic to the ring of symmetric polynomials with $k$ variables,
which are just the diagonal elements of the $\sigma$ field, as
described in section~\ref{sect:gauge-inv-ops}. Let's denote this
ring by $A(k)$.\par

Because the connecting map
$$\delta: \:
A(k) \: \longrightarrow \: \underset{r\geqslant 0}{\oplus} H^r(\
\wedge^r \mathcal{E}^*)$$ (given as the sum of all $\delta_r$'s) is
a surjective homomorphism of $\mathbb{C}$-algebras as proved in
\cite{3}, to determine the ring structure of the latter, we only
need to find $\ker(\delta )$, the kernel of the connecting map. Thus
we see, the sheaf cohomology has the following representation
\begin{equation}
\underset{r\geqslant 0}{\oplus} H^r\left(G(k,n), \wedge^r
\mathcal{E}^*\right) \cong A(k)/\ker(\delta).
\end{equation} \par

As in the text, we shall denote by $\sigma_{(r)}$ the Schur
polynomial corresponding to the Young diagram with $r$ boxes in a
row. In \cite{3}, we have shown that, for a generic $B$ deformation,
$\ker(\delta)$ is generated by $R_{(r)}, r=n-k+1, \cdots,$ where
\begin{equation}
R_{(r)} = \sum_{i=0}^{\min\{r,n\}} I_i \sigma_{(r-i)}
\sigma_{(1)}^i,
\end{equation}
and $I_i$ is the $i$th characteristic polynomial of $B$, which is
defined through
\[
\det(t I + B) = \sum_{i=0}^n I_{n-i} t^i.
\]
This gives us the classical sheaf cohomology. (In
appendix~\ref{app:trivial-defs}, we give a toy model of the
mathematical arguments for general cases that will appear in
\cite{3}, and mathematically derive ring relations for trivial
bundle deformations $B \propto I$.) In section \ref{derivation}, we
derive the quantum sheaf cohomology for generic deformations from
the one-loop effective action. The conclusion is, when quantum
corrections are taken into account, we should replace $\ker(\delta)$
with the ideal generated by
\begin{equation}\label{q-R}
R_{(r)} + q \sigma_{(r-n)},
\end{equation}
$r=n-k+1,\cdots$, and $\sigma_{(m)}$ is defined to be zero when
$m<0$.

In section~\ref{sect:qsc-ring-details}, we describe the quantum
sheaf cohomology ring explicitly.  Since
\begin{displaymath}
R_{\mu} \: \in \: \left\langle R_{(n-k+1)}, R_{(n-k+2)}, \cdots
\right\rangle,
\end{displaymath}
for Young diagrams $\mu$ with more than $n-k$ boxes in the first row
and no more than $k$ boxes in any column, our description of the
ring structure suggests that the space
\[
W_m = \mathrm { Span }_\mathbb{C} \left\{ \, R_{\mu} \,  | \, \mu_1
> n-k , | \mu | = m \right\}
\]
is in $\ker(\delta_m)$ for each $m > n-k$.

If the $R_{\mu}$'s are linearly independent, we see
$$\dim W_m = \dim
\ker (\left. \delta_m \right| _{B=0}),$$ thus $W_m =
\ker(\delta_m)$. Our description is complete if this is true for all
$m > n-k$.

If, on the other hand, there is some $m$, such that the $R_{\mu}$'s
are linearly dependent, then $W_m$ is only a proper subspace of
$\ker(\delta_m)$. This situation occurs along a locus $V_m$ with
codimension at least one in the moduli space. If we expand $R_\mu$
in terms of $\sigma_\nu$, we get
\[
R_\mu = \sum_{| \nu | = m} C^m_{\mu \nu} \, \sigma_\nu
\]
with some constants $C^m_{\mu \nu}$ depending on $B$. For fixed $m$,
$C^m_{\mu \nu}$ form a matrix. So by definition, $V_m$ is the common
zero locus of all the $p$ minors of this matrix, where $p$ is the
number of Young tableaux whose first row has more than $n-k$ boxes
and with no more than $k$ boxes in any column. Thus, along this
locus, the dimension of the ring, and the dimensions of the
corresponding sheaf cohomology groups, might jump, so we exclude
this locus from our discussion of the quantum sheaf cohomology ring.

Now, since $V_m$ is defined by the kernel of $\delta_m$, which maps
into degree $m$ cohomology of $G(k,n)$, the dimensions of the sheaf
cohomology groups we are interested in can potentially jump along
the loci $V_m$ for
\begin{displaymath}
n-k+1 \: \leq \: m \: \leq \: k(n-k),
\end{displaymath}
as there is no cohomology of degree greater than the dimension of
the Grassmannian. Along loci $V_m$ for $m > k(n-k)$, our description
of the quantum sheaf cohomology ring may be incomplete, but the
dimensions of the sheaf cohomology groups cannot jump.

As a practical matter, we can phrase this in terms of two different
representations of the classical sheaf cohomology ring, as follows.
One way to present the classical sheaf cohomology ring is
\begin{displaymath}
{\mathbb C}\left[ \sigma_{(1)}, \cdots, \sigma_{(k(n-k))} \right] /
\left\langle D_{k+1}, \cdots, R_{(n-k+1)}, \cdots, R_{(k(n-k))},
S_{k(n-k)+1}(k) \right\rangle,
\end{displaymath}
where $S_r(k)$ is the ideal of degree $r$ terms in $k$ variables.
This presentation of the classical sheaf cohomology ring is valid on
\begin{displaymath}
{\cal M} - \left( X \cup V_{n-k+1} \cup \cdots \cup V_{k(n-k)}
\right)
\end{displaymath}
where ${\cal M}$ is the space of all $B$ deformations, $X$ is the
discriminant locus along which any $k$ eigenvalues of $B$ become
$-1$ (equivalent to the locus $\{ \Delta = 0 \}$ for the $\Delta$
appearing explicitly in correlation functions), and $V_r$ is defined
as above. Along the $V_r$ for the degrees above, the dimensions of
the sheaf cohomology groups may potentially jump.

A second presentation of the same classical sheaf cohomology ring is
\begin{displaymath}
{\mathbb C}\left[ \sigma_{(1)}, \sigma_{(2)}, \cdots \right] /
\left\langle D_{k+1}, D_{k+2}, \cdots, R_{(n-k+1)}, R_{(n-k+2)},
\cdots \right\rangle,
\end{displaymath}
which is closer to the form we have adopted in these notes for the
quantum sheaf cohomology ring.  This presentation is valid on
\begin{displaymath}
{\cal M} - \left( X \cup V_{n-k+1} \cup \cdots \cup \cdots \right).
\end{displaymath}
In this case, the $V_r$ for
\begin{displaymath}
n-k+1 \: \leq \: r \: \leq \: k(n-k)
\end{displaymath}
define loci along which the classical sheaf cohomology groups might
jump, and the $V_r$ for $r > k(n-k)$ define loci along which the
dimensions of the sheaf cohomology groups cannot jump, but, for
which the presentation above may not be accurate, as additional
relations may be required.

\section{Products via homological algebra}
\label{app:hom-alg}

It is possible to see at least the classical product structure on
the (2,2) locus in homological algebra, by identifying sheaf
cohomology groups with Ext groups and interpreting in terms of
extensions of bundles, an idea also discussed in \cite{dgks1,dgks2}.
In this appendix, we will look at that structure in simple cases,
first describing how products in the ordinary cohomology of the
Grassmannian can be computed via homological algebra, and later
outlining some of the machinery needed to do analogous computations
in the (0,2) case. We will not use this machinery elsewhere in this
paper, but we felt it sufficiently interesting to include here as an
appendix. Furthermore, because of its existence, we speculate that
perhaps `quantum sheaf cohomology' can be understood as a `quantum
homological algebra'.

Let us begin on the (2,2) locus, and describe products in the
ordinary cohomology of a Grassmannian via homological algebra. On
the (2,2) locus, as is well-known, the cohomology of $G(k,n)$ is
generated by Chern classes of the (dual of the) universal subbundle
$S$. In that spirit, we can identify $c_1$ with an element of
Ext$^1(S^*,Q^*)$, corresponding to the complex
\begin{displaymath}
0 \: \longrightarrow \: Q^* \: \longrightarrow \: V^*
\end{displaymath}
resolving $S^*$, where $Q$ is the universal quotient bundle and $V$
is a vector space, $G(k,n) = G(k,V)$. As a consistency check, note
that there is a natural map
\begin{displaymath}
{\rm Ext}^1(S^*, Q^*) \: \longrightarrow \: H^1(S\otimes Q^* \otimes
S \otimes S^* \otimes Q \otimes Q^*) \: = \: H^1\left(\Omega^1
\otimes {\rm End}\, T\right),
\end{displaymath}
which displays how to map Ext$^1(S^*,Q^*)$ to the sheaf cohomology
group containing the Atiyah class of the tangent bundle, and is
ultimately the reason that the identification of $c_1$ with an
element of the Ext group above is sensible.

Recall from section~\ref{sect:gauge-inv-ops} the product
\begin{displaymath}
\sigma_{\tiny\yng(1)}^2 \: = \: \sigma_{\tiny\yng(2)} \: + \:
\sigma_{\tiny\yng(1,1)}.
\end{displaymath}
We can understand this in the present language as follows. The
product $\sigma_{\tiny\yng(1)}^2$ should be understood as
\begin{displaymath}
(c_1)^2 \: \in \: {\rm Ext}^2\left( S^* \otimes S^*, Q^* \otimes Q^*
\right),
\end{displaymath}
corresponding to the resolution of $S^* \otimes S^*$ below,
\begin{displaymath}
0 \: \longrightarrow \: Q^* \otimes Q^* \: \longrightarrow \: Q^*
\otimes V^* + Q^* \otimes V^* \: \longrightarrow \: V^* \otimes V^*
\: \longrightarrow \: S^* \otimes S^* \: \longrightarrow \: 0,
\end{displaymath}
given by squaring the resolution of $S^*$. This naturally decomposes
into the sum of the following two resolutions:
\begin{equation} \label{eq:sigma2}
0 \: \longrightarrow \: \wedge^2 Q^* \: \longrightarrow \: Q^*
\otimes V^* \: \longrightarrow \: {\rm Sym}^2 V^* \: \longrightarrow
\: {\rm Sym}^2 S^* \: \longrightarrow \: 0,
\end{equation}
and
\begin{equation} \label{eq:sigma11}
0 \: \longrightarrow \: {\rm Sym}^2 Q^* \: \longrightarrow \: Q^*
\otimes V^* \: \longrightarrow \: \wedge^2 V^* \: \longrightarrow \:
\wedge^2 S^* \: \longrightarrow \: 0.
\end{equation}
If we identify the resolution~(\ref{eq:sigma2}) with
\begin{displaymath}
\sigma_{\tiny\yng(2)} \: \in \: {\rm Ext}^2\left( {\rm Sym}^2 S^*,
\wedge^2 Q^* \right),
\end{displaymath}
and the resolution~(\ref{eq:sigma11}) with
\begin{displaymath}
\sigma_{\tiny\yng(1,1)} \: \in \: {\rm Ext}^2\left( \wedge^2 S^*,
{\rm Sym}^2 Q^* \right),
\end{displaymath}
then we recover
\begin{displaymath}
\sigma_{\tiny\yng(1)}^2 \: = \: \sigma_{\tiny\yng(2)} \: + \:
\sigma_{\tiny\yng(1,1)}.
\end{displaymath}

As a consistency check, note that in the language of Atiyah classes,
the cohomology classes above should live in
\begin{displaymath}
H^2\left( \Omega^2 \otimes {\rm End}\, T\right) \: = \: H^2\left(
\wedge^2(S \otimes Q^*) \otimes S \otimes S^* \otimes Q \otimes Q^*
\right),
\end{displaymath}
and as
\begin{displaymath}
\wedge^2(S \otimes Q^*) \: = \: \wedge^2 S \otimes {\rm Sym}^2 Q^*
\: + \: {\rm Sym}^2 S \otimes \wedge^2 Q^*,
\end{displaymath}
we see that both $\sigma_{\tiny\yng(2)}$ and
$\sigma_{\tiny\yng(1,1)}$ naturally map into $H^2(\Omega^2 \otimes
{\rm End}\,T)$, as expected.

Now, let us turn to the (0,2) case.  Here, we have a deformation
${\cal E}$ of the tangent bundle defined by a short exact sequence
\begin{displaymath}
0 \: \longrightarrow \: {\cal E}^* \: \longrightarrow \: V^* \otimes
S \: \longrightarrow \: S^* \otimes S \: \longrightarrow \: 0.
\end{displaymath}
Applying Hom$({\cal O},-)$ to the sequence above, one gets
\begin{displaymath}
0\to {\rm Hom}(\mc{O}, \mc{S}^*\otimes\mc{S}) \xrightarrow{\delta}
{\rm Ext}^1(\mc{O},\mc{E}^*).
\end{displaymath}

In the rest of this section, we will begin to outline some of the
machinery needed to compute classical sheaf cohomology rings via
homological algebra.

Note that from the construction of the connecting morphism, one
finds that\footnote{See the remark after Theorem III.5.2 of
\cite{MR1438546}.} for any $\varphi \in {\rm
Hom}(\mc{O},\mc{S}^*\otimes\mc{S})$, one has $\delta(\varphi) =
[E_\varphi]\in {\rm Ext}^1(\mc{O},\mc{E}^*)$, where
\begin{displaymath}
E_\varphi: 0\to\mc{E}^* \to \mc{Z}  \to \mc{O} \to 0
\end{displaymath}
is constructed from the pullback diagram
\begin{displaymath}
\xymatrix{
\mc{Z}\ar[d]\ar[r]&\mc{O} \ar[d]^{\varphi}\\
\mc{V}^*\otimes\mc{S}\ar[r]& \mc{S}^*\otimes\mc{S} },
\end{displaymath}
which fits in
\begin{displaymath}
\xymatrix{
0\ar[r]&\mc{E}^*\ar@{=}[d]\ar[r]& \mc{Z}\ar[d]\ar[r]&\mc{O} \ar[d]^{\varphi}\ar[r]&0\\
0\ar[r]&\mc{E}^*\ar[r]& \mc{V}^*\otimes\mc{S}\ar[r]&
\mc{S}^*\otimes\mc{S}\ar[r]&0 }.
\end{displaymath}

Applying the functor $\Sym^2$ to
$\mc{O}\xrightarrow{\varphi}\mc{S}^*\otimes\mc{S}$, we get the
induced map
$$\mc{O}\xrightarrow{\Sym^2\varphi}\Sym^2 (\mc{S}^*\otimes\mc{S}).$$
It is easy to construct the induced extension sequence following the
method in last section, and there is no difficulty to do similar
things for any map $\mc{O}\to \Sym^r(\mc{S}^*\otimes\mc{S})$.

Now denote the image of $Id: K_\lambda \mc{S} \to K_\lambda\mc{S}$
in $H^0(K_\lambda\mc{S}^*\otimes K_\lambda\mc{S})$ and
Hom$(\mc{O},K_\lambda\mc{S}^*\otimes K_\lambda\mc{S})$ by
$\kappa_\lambda$. We strongly suspect, but have not carefully
checked, that
\begin{itemize}
\item The multiplication $\kappa_1\cdot\kappa_1\cdot\cdots\cdot \kappa_1 := \Sym^r \kappa_1$ agrees with the ring structure of $H^\bullet(\wedge^\bullet \mc{E}^*)$.
\item The multiplication
$\kappa_\alpha \cdot \kappa_\beta =
\Phi(\kappa_\alpha\otimes\kappa_\beta)$ agrees with the ring
structure of $H^\bullet(\wedge^\bullet \mc{E}^*)$, where $\Phi$ is
the symmetrization map
$$\Sym^{|\alpha|}(\mc{S}^*\otimes\mc{S})\otimes\Sym^{|\beta|}(\mc{S}^*
\otimes\mc{S})\to \Sym^{|\alpha|+|\beta|}(\mc{S}^*\otimes\mc{S}).$$
\end{itemize}

Now, let us sketch out some computations in the case $n=2$.

Let $\varphi_1,\varphi_2 \in {\rm Hom}(\mc{O},\mc{S^*\otimes S})$,
$\varphi =\varphi_1\otimes \varphi_2 \in
 {\rm Hom}(\mc{O},\Sym^2(\mc{S^*\otimes S}))$.
We want to compare $E_{\varphi_1}\cdot E_{\varphi_2}$ with
$E^2_\varphi \in {\rm Ext}^2(\mc{O},\wedge^2\mc{E}^*)$.

Note that for $i=1,2$ we have
\begin{displaymath}
 E_{\varphi_i}:0 \to \mc{E}^* \to F_i \to \mc{O} \to 0
\end{displaymath}
and tensoring with $\mc{E}^*$, we have
\begin{displaymath}
0 \to \mc{E}^*\otimes \mc{E}^* \to F_2\otimes \mc{E}^* \to \mc{E}^*
\to 0.
\end{displaymath}
This gives
\begin{displaymath}
 \xymatrix{
&0\ar[r]&\mc{E}^*\otimes\mc{E}^*\ar[d]\ar[r]& F_2\otimes \mc{E}^*\ar[d]\ar[r]&\mc{E}^* \ar@{=}[d]\ar[r]&0\\
\tilde{E}_{\varphi_2}:&0\ar[r]&\wedge^2\mc{E}^*\ar[r]&
\tilde{F}_2\ar[r]& \mc{E}^*\ar[r]&0 },
\end{displaymath}
where the first square is a push-out diagram defining $\tilde{F}_2$.

So we can define $E_{\varphi_1}\cdot E_{\varphi_2}$ as
$E_{\varphi_1}\cdot \tilde{E}_{\varphi_2}$ which is represented by
\begin{displaymath}
0 \to \wedge^2\mc{E}^* \to \tilde{F}_{_2} \to F_1 \to \mc{O}\to 0.
\end{displaymath}
On the other hand, for $\varphi$ we have $E_\varphi$ represented by
\begin{displaymath}
0 \to \wedge^2 \mc{E}^* \to \wedge^2(\mc{V^*\otimes S}) \to F \to
\mc{O}\to 0.
\end{displaymath}

We claim that
 $[E_{\varphi_1}\cdot E_{\varphi_2}] =[E_\varphi]$ in
Ext$^2(\mc{O},\wedge^2\mc{E}^*)$.

We can see this as follows. Recall that two $n$-extensions $H_0$ and
$H_m$ are equivalent iff they are connected by some morphisms of
complexes: $H_0 \to H_1 \leftarrow H_2 ... \leftarrow H_m$, where
each $H_a\to H_b$, $(a,b)=(i,i+1)$ or $(a,b)=(i+1,i)$ is of the form
\begin{equation}\label{CD_n_extension}
\xymatrix{
0 \ar[r] & B \ar[r]\ar@{=}[d] & E_n^a \ar[r]\ar[d] &... \ar[r] & E_1^a \ar[r]\ar[d] & A \ar[r]\ar@{=}[d] & 0\\
0 \ar[r] & B \ar[r] & E_n^b \ar[r] &... \ar[r] & E_1^b \ar[r] & A
\ar[r] & 0 &. }
\end{equation}

Let $\mc{Z}=\mc{V^*\otimes S}$ and
$\mc{E}nd=End(\mc{S})=\mc{S^*\otimes S}$. We need to show that there
exist maps $\alpha$ and $\beta$ such that all squares in the
following diagram are commutative:
{
\newcommand{\eAA}{\mc{E}^*\otimes\mc{E}^*}
\newcommand{\eAB}{F_2\otimes\mc{E}^*}
\newcommand{\eAC}{F_1}
\newcommand{\eAD}{\mc{O}}
\newcommand{\eBA}{\wedge^2\mc{E}^*}
\newcommand{\eBB}{\tilde{F}_2}
\newcommand{\eBC}{F_1}
\newcommand{\eBD}{\mc{O}}
\newcommand{\eCA}{\wedge^2\mc{E}^*}
\newcommand{\eCB}{\wedge^2\mc{Z}}
\newcommand{\eCC}{F}
\newcommand{\eCD}{\mc{O}}
\newcommand{\eDA}{\wedge^2\mc{E}^*}
\newcommand{\eDB}{\wedge^2\mc{Z}}
\newcommand{\eDC}{\mc{Z}\otimes\mc{E}nd}
\newcommand{\eDD}{\Sym^2\mc{E}nd}
%
%
\begin{equation}\label{4x4}
\xymatrix{
0\ar[r] &       \eAA\ar[r]\ar[d]        &       \eAB\ar[r]\ar[d]        &       \eAC\ar[r]\ar@{=}[d]    &       \eAD\ar[r]\ar@{=}[d]    &0\\
0\ar[r] &       \eBA\ar[r]\ar@{=}[d]    &       \eBB\ar[r]\ar@{-->}[d]^{\beta}  &       \eBC\ar[r]^{p_1}\ar@{-->}[d]^{\alpha}   &       \eBD\ar[r]\ar@{=}[d]    &0\\
0\ar[r] &       \eCA\ar[r]\ar@{=}[d]    &       \eCB\ar[r]\ar@{=}[d]    &       \eCC\ar[r]^{p_2}\ar[d]  &       \eCD\ar[r]\ar[d]        &0\\
0\ar[r] &       \eDA\ar[r]      &       \eDB\ar[r]      &
\eDC\ar[r]      &       \eDD\ar[r]      &0 }
\end{equation}
For $\alpha$, notice that we have a map $\alpha_0: F_1 \to \eDC$
constructed as in
\begin{displaymath}
\xymatrix{
\eBC \ar[r]\ar[d]                               &\eBD\ar[d]^{\varphi_1\otimes 1} \\
\mc{Z}\otimes \mc{O}\ar[r]\ar[d]        &\mc{E}nd\otimes\mc{O}\ar[d]^{1\otimes\varphi_2}\\
\mc{Z}\otimes \mc{E}nd\ar[r]\ar@{=}[d]  &\mc{E}nd\otimes\mc{E}nd\ar[d]^{s}&.\\
\mc{Z}\otimes \mc{E}nd\ar[r]    &\Sym^2\mc{E}nd }
\end{displaymath}

Also, the composition of the maps of second column is exactly
$\varphi$. Since the square at the lower right corner of (\ref{4x4})
is a pull-back square, there exists $\alpha$ such that $\alpha_0$
factors through $\alpha$.

Similarly, we can construct $\eAB\to \eCB$ canonically via:
\begin{displaymath}
\xymatrix{
\eAA \ar[r]\ar@{=}[d]                   & \eAB \ar[d]\\
\eAA \ar[r]\ar@{=}[d]                   & \mc{Z}\otimes\mc{E}^*\ar[d]\\
\eAA \ar[r]\ar[d]                       & \mc{Z}\otimes\mc{Z}\ar[d] &.\\
\eCA \ar[r]                                     & \eCB }
\end{displaymath}
Since the square at the upper left corner of (\ref{4x4}) is a
push-forward square, there exists $\beta$ such that $\beta_0$
factors through $\beta$.

Now it suffices to prove that the central square of (\ref{4x4}) is
commutative. But we have already know that the second and third rows
of (\ref{4x4}) are exact. So the commutativity of the central square
is the consequence of that of

$$\xymatrix{
\eBB\ar[r]\ar[d] & S^1_1\ar[d] \\
\eCB\ar[r]              & S^2_1 }\rm{\ \ \ and\ \ \ } \xymatrix{
S^1_1\ar[r]\ar[d] & \eBC\ar[d] \\
S^2_1\ar[r]             & \eCC },$$ where $S_1^i = Ker\ p_i, i=1,2$.
}

Next, we turn to general $n$.

We want to prove the following diagram is commutative:
\begin{displaymath}
\xymatrix{
( H^0(\Sym^r(\mc{E}nd)), H^0(\Sym^r(\mc{E}nd)) ) \ar[r]\ar[d]   & H^0(\Sym^{r+s}(\mc{E}nd))\ar[d]\\
( H^r(\wedge^r\mc{E}^*), H^s(\wedge^s\mc{E}^*) ) \ar[r]         &
H^{r+s}(\wedge^{r+s}\mc{E}^*) } .
\end{displaymath}
We do so by identifying it with
\begin{equation}\label{ext_multi}
\xymatrix{
( {\rm Hom}(\mc{O},\Sym^r(\mc{E}nd)), {\rm Hom}(\mc{O},\Sym^s(\mc{E}nd)) ) \ar[r]\ar[d]     & {\rm Hom}(\mc{O},\Sym^{r+s}(\mc{E}nd))\ar[d]\\
( {\rm Ext}^r(\mc{O},\wedge^r\mc{E}^*), {\rm
Ext}^s(\mc{O},\wedge^s\mc{E}^*) ) \ar[r]               & {\rm
Ext}^{r+s}(\mc{O}, \wedge^{r+s}\mc{E}^*) } .
\end{equation}

We claim that the diagram~(\ref{ext_multi}) is commutative. We can
see this as follows. Take $\varphi_1\in {\rm
Hom}(\mc{O},\Sym^r(\mc{E}nd))$, $\varphi_2\in {\rm
Hom}(\mc{O},\Sym^s(\mc{E}nd))$. Define the multiplication
\begin{displaymath}
\begin{array}{clc}
{\rm Hom}(\mc{O},\Sym^r(\mc{E}nd)) \times {\rm
Hom}(\mc{O},\Sym^s(\mc{E}nd))
 &\to & {\rm Hom}(\mc{O},\Sym^{r+s}(\mc{E}nd))\\
(\varphi_1, \varphi_2)&\mapsto &\varphi
\end{array},
\end{displaymath}
where $\varphi$ is defined by the composition
\begin{displaymath}
\mc{O}\xrightarrow{\varphi_1\otimes 1} \Sym^r(\mc{E}nd)\otimes\mc{O}
\xrightarrow{1\otimes \varphi_2} \Sym^r(\mc{E}nd)\otimes
\Sym^s(\mc{E}nd) \xrightarrow{\text{symmetrize}}
\Sym^{r+s}(\mc{E}nd).
\end{displaymath}
Up to a minus sign, we have a sequence $E_{\varphi_2}$ representing
$\Delta(\varphi_2)\in {\rm Ext}^{s}(\mc{O},\wedge^{s}\mc{E}^*)$
\begin{equation}\label{sym_r_to_ext}
\xymatrix{
E_{\varphi_2}:&0\ar[r]  & \wedge^{s}\mc{E}^* \ar[r]\ar@{=}[d]   & ... \ar[r]    & F_2 \ar[r]\ar[d]      & \mc{O} \ar[r]\ar[d]^{\varphi_2} &0\\
&  0\ar[r]      & \wedge^{s}\mc{E}^* \ar[r]     & ... \ar[r]    &
\mc{Z}\otimes \Sym^{s-1}\mc{E}nd \ar[r]       & \Sym^{s}\mc{E}nd
\ar[r] &0}
\end{equation}
and similarly we have $E_{\varphi_1}$ and $E_{\varphi}$ for
$\Delta(\varphi_1)$ and $\Delta(\varphi)$, with the pull-back sheaf
$F_2$ replaced by $F_1$ and $F$ respectively.

Tensoring (\ref{sym_r_to_ext}) with $\wedge^r\mc{E}^*$, we get
\begin{equation}\label{tensored_wedge_r}
\xymatrix{ \ \ \ \ \ \ \ 0\ar[r]   &
\wedge^{s}\mc{E}^*\otimes\wedge^r\mc{E}^* \ar[r]\ar[d]
& \wedge^{s}\mc{Z}\otimes\wedge^r\mc{E}^* \ar[r]\ar[d]  & ... \ar[r]    & F_2\otimes\wedge^r\mc{E}^* \ar[r]\ar@{=}[d]   & \wedge^r\mc{E}^* \ar[r]\ar@{=}[d] &0\\
\tilde{E}_{\varphi_1}:  0\ar[r]         & \wedge^{r+s}\mc{E}^*
\ar[r]   & \tilde{F}_2\ar[r] &... \ar[r]         &
F_2\otimes\wedge^r\mc{E}^* \ar[r]     & \wedge^r\mc{E}^* \ar[r] &0}
\end{equation}
where the first square is a push-out diagram.

To verify  $[E_{\varphi_1}\cdot E_{\varphi_2}] =[E_\varphi]$ in
Ext$^{r+s}(\mc{O},\wedge^{r+s}\mc{E}^*)$, we break the sequence
$E_\varphi$ into
\begin{displaymath}
0\to S_r \to ...\to \mc{Z}\otimes\Sym^{r+s-1}\mc{E}nd \to
\Sym^{r+s}\mc{E}nd \to 0
\end{displaymath}
and
\begin{displaymath}
0\to\wedge^{r+s}\mc{E}^*\to\wedge^{r+s}\mc{Z}\to...\to S_r\to 0.
\end{displaymath}

As a first step, we show that squares in the last two rows of the
following diagram are commutative:
{
\newcommand{\eAA}{\wedge^s\mc{E}^*\otimes\wedge^r\mc{E}^*}
\newcommand{\eAB}{\wedge^{s}\mc{Z}\otimes\wedge^r\mc{E}^*}
\newcommand{\eAC}{F_2\otimes\wedge^r\mc{E}^*}
\newcommand{\eAD}{\wedge^r\mc{E}^*}
\newcommand{\eBA}{\wedge^{r+s}\mc{E}^*}
\newcommand{\eBB}{\tilde{F}_2}
\newcommand{\eBC}{\eAC}
\newcommand{\eBD}{\eAD}
\newcommand{\eDA}{\wedge^{r+s}\mc{E}^*}
\newcommand{\eDB}{\wedge^{r+s}\mc{Z}}
\newcommand{\eDC}{\wedge^{r+1}\mc{Z}\otimes \Sym^{s-1}\mc{E}nd}
\newcommand{\eDD}{S_r}
%
%
\begin{displaymath}
\xymatrix{
0\ar@{->}[r]    &       \eAA\ar[r]\ar[d]        &       \eAB\ar[r]\ar[d]        & ...\ar[r]&    \eAC\ar[r]\ar@{=}[d]    &       \eAD\ar[r]\ar@{=}[d]    &0\\
0\ar[r] &       \eBA\ar[r]\ar@{=}[d]    &       \eBB\ar[r]\ar@{-->}[d]^{\beta_s}        & ...\ar[r]&    \eBC\ar[r]\ar@{-->}[d]^{\beta_1}        &       \eBD\ar[r]\ar@{-->}[d]^{\beta_0}        &0\\
0\ar[r] &       \eDA\ar[r]      &       \eDB\ar[r]      & ...\ar[r]&
\eDC\ar[r]      &       \eDD\ar[r]      &0 }
\end{displaymath}

To see this, we first need to define the maps $\beta_j$,
$j=0,...,s$. Since the upper left square is a push-out square, to
define $\beta_s$ it suffices to find a map $\beta_{s,0}$ such that
$$\xymatrix{
\eAA\ar[r]\ar[d] &\eAB\ar[d]^{\beta_{s,0}}\\
\eDA\ar[r] &\eDB }$$ commutes. But we have a canonical choice of
$\beta_{s,0}$, namely
$$\eAB\to \wedge^s\mc{Z}\otimes\wedge^r\mc{Z}\to\eDB$$
with obvious maps. For $j=s-1, s-2,...,2$, $\beta_{j}$ is the
canonical map
\begin{displaymath}
\begin{array}{clc}
\wedge^j\mc{Z}\otimes\wedge^r\mc{E}^*\otimes \Sym^{s-j}\mc{E}nd &\to & \wedge^{r+j}\mc{Z}\otimes \Sym^{s-j}\mc{E}nd\\
u \otimes v \otimes w & \mapsto & u \wedge v \otimes w
\end{array}
\end{displaymath}
 and it is easy to see that squares with vertical maps in $\{\beta_j|j=s-1,s-2,...,2\}$ are commutative.

To see that the square containing $\beta_s$ and $\beta_{s-1}$ is
commutative, we use the diagram
\begin{equation}\label{CD_beta}
\xymatrix{
0\ar@{->}[r]    &       \eAA\ar[r]\ar[d]^{\beta_{s+1}^\prime}   &       \eAB\ar[r]^{d_s\ \ \ \ }\ar[d]^{\beta_s^\prime} & \wedge^{s-1}\mc{Z}\otimes\wedge^r\mc{E}^*\otimes \mc{E}nd \ar@{=}[d]^{\beta_{s-1}^\prime}\\
0\ar[r] &       \eBA\ar[r]\ar@{=}[d]    &       \eBB\ar[r]^{\tilde{d}_s\ \ \ \ }\ar@{->}[d]^{\beta_s}   & \wedge^{s-1}\mc{Z}\otimes\wedge^r\mc{E}^*\otimes \mc{E}nd \ar[d]^{\beta_{s-1}}\\
0\ar[r] &       \eDA\ar[r]      &       \eDB\ar[r]^{D_s\ \ \ \ }
& \wedge^{r+s-1}\mc{Z}\otimes \mc{E}nd}\ .
\end{equation}
By the commutativity of the lower right square and the square
containing $d_s$ and $D_s$, $\beta_{s-1}\circ
\tilde{d}_s\circ\beta_s^\prime =\beta_{s-1}\circ
\beta_{s-1}^\prime\circ d_s = D_s\circ\beta_s\circ\beta_s^\prime$.

Since $\tilde{F}_2$ is a push-out, and $\beta_{s+1}^\prime$ is
surjective, so is $\tilde{\beta}_s^\prime$. Hence we have
$\beta_{s-1}\circ \tilde{d}_s = D_s\circ\beta_s$. Hence the square
containing $\beta_s$ and $\beta_{s-1}$ is commutative.

Now we define $\beta_1, \beta_0$. They are canonically defined by
the second and third columns of the following diagram:

\begin{displaymath}
\xymatrix{ \wedge^2\mc{Z}\otimes \Sym^{s-2}\mc{E}nd\otimes
\wedge^r\mc{E}^*        \ar[r]\ar[d]            & \eAC \ar[r]\ar[d]
& \eAD \ar[d]
\\
\wedge^2\mc{Z}\otimes \Sym^{s-2}\mc{E}nd\otimes \wedge^r\mc{E}^*        \ar[r]\ar[d]            &\mc{Z}\otimes \Sym^{s-1}\mc{E}nd\otimes \wedge^r\mc{E}^* \ar[r]\ar[d]  & \Sym^s\mc{E}nd\otimes \wedge^r{E}^*   \ar[d]\\
\wedge^{r+2}\mc{Z}\otimes\Sym^{s-2}\mc{E}nd \ar[r]
&\wedge^{r+1}\mc{Z}\otimes\Sym^{s-1}\mc{E}nd  \ar[r]
&\wedge^{r}\mc{Z}\otimes\Sym^{s}\mc{E}nd . }
\end{displaymath}
Commutativity of squares involving $\beta_1, \beta_0$ are obvious. }

The next step is to show that squares in the first two rows of the
following diagram are commutative:

{
\newcommand{\eAA}{\wedge^r\mc{E}^*\otimes \mc{O}}
\newcommand{\eAB}{\wedge^{r}\mc{Z}\otimes \mc{O}}
\newcommand{\eAC}{F_1\otimes \mc{O}}
\newcommand{\eAD}{\mc{O}}
\newcommand{\eCA}{S_r}
\newcommand{\eCB}{\eDB}
\newcommand{\eCC}{F}
\newcommand{\eCD}{\mc{O}}
\newcommand{\eDA}{S_r}
\newcommand{\eDB}{\wedge^{r}\mc{Z}\otimes\Sym^{s}\mc{E}nd}
\newcommand{\eDC}{\mc{Z}\otimes \Sym^{r+s-1} \mc{E}nd}
\newcommand{\eDD}{\Sym^{r+s}\mc{E}nd}
%
%
\begin{displaymath}
\xymatrix{
        \eAA\ar@{^{(}->}[r]\ar[d]       ^{\beta_0}&     \eAB\ar[r]\ar@{-->}[d]  ^{\alpha_r}& ...\ar[r]& \eAC\ar@{->>}[r]\ar@{-->}[d]^{\alpha_1} &       \eAD\ar@{-->}@{=}[d]\\
\eCA\ar@{^{(}->}[r]\ar@{=}[d]   &       \eCB\ar[r]\ar@{=}[d]    & ...\ar[r]&    \eCC\ar@{->>}[r]\ar@{->}[d]^{\alpha_1^\prime}   &       \eCD\ar@{->}[d]^{\varphi}\\
\eDA\ar@{^{(}->}[r]     &       \eDB\ar[r]      & ...\ar[r]&
\eDC\ar@{->>}[r]        &       \eDD }.
\end{displaymath}
For $j=r,r-1,...,2$, $\alpha_j$ is defined by
$(\wedge^j\mc{Z}\otimes\Sym^{r-j}\mc{E}nd)\otimes\mc{O}
\xrightarrow{1\otimes \varphi_2}
(\wedge^j\mc{Z}\otimes\Sym^{r-j}\mc{E}nd)\otimes\Sym^{s}\mc{E}nd
\xrightarrow{\text{symmetrization}}
\wedge^j\mc{Z}\otimes\Sym^{r+s-j}\mc{E}nd$, where symmetrization is
the restriction of the projection $\mc{E}nd^{\otimes r+s-j}\to
\Sym^{r+s-j}\mc{E}nd$ to $\Sym^{r-j}\mc{E}nd\otimes\Sym^{s}\mc{E}nd
\subset \mc{E}nd^{\otimes r+s-j}$. The squares involving them are
obviously commutative.

The following diagram and the universal property of $F$ defines
$\alpha_1$.

\begin{displaymath}
\xymatrix{
F_1 \otimes \mc{O} \ar[r]\ar[d]                                 &\mc{O} \otimes \mc{O} \ar[d]^{\varphi_1\otimes 1} \\
\mc{Z}\otimes \Sym^{r-1} \mc{E}nd \ar[r]\ar[d]  &\Sym^r \mc{E}nd\otimes\mc{O}\ar[d]^{1\otimes\varphi_2}\\
\mc{Z}\otimes \Sym^{r-1} \mc{E}nd \otimes \Sym^{s} \mc{E}nd \ar[r]\ar[d]        & \Sym^{r} \mc{E}nd \otimes \Sym^{s} \mc{E}nd\ar[d]&.\\
\mc{Z}\otimes \Sym^{r+s-1} \mc{E}nd\ar[r]       &\Sym^{r+s}\mc{E}nd
}
\end{displaymath}
Then an argument dual to the one regarding (\ref{CD_beta}) shows
that the square containing $\alpha_1$ and $\alpha_2$ is commutative.
}

Combining the two steps above, we have constructed a morphism of
complexes, which shows the desired equality of extension classes in
view of (\ref{CD_n_extension}).

\section{Some identities for trivial deformations}
\label{app:trivial-defs}

In this section we will mathematically derive ring relations in the
special case of a trivial (0,2) deformation defined by a nonzero $B
\propto I$.  For reasons discussed in section~\ref{sect:basicphys},
this deformation does not change the tangent bundle.  This is, in
essence, a toy model of the mathematical arguments for general cases
that will appear in \cite{3}.

The idea is to run the relations in the ordinary cohomology ring of
the Grassmannian, through the isomorphism between the `deformed'
tangent bundle and the standard presentation of the tangent bundle,
to get a prototype for the classical sheaf cohomology ring relations
in more general cases.

To this end, we define a map
\begin{displaymath}
h: \: S^* \otimes S \: \longrightarrow \: S^* \otimes S
\end{displaymath}
that makes the diagram
\begin{displaymath}
\xymatrix{ 0 \ar[r] & S^* \otimes S \ar[r] \ar[d]_h & S^* \otimes V
\ar[r] \ar@{=}[d]
& {\cal E} \ar[r] \ar[d]_{\cong} & 0 \\
0 \ar[r] & S^* \otimes S \ar[r] & S^* \otimes V \ar[r] & T \ar[r] &
0 }
\end{displaymath}
commute, where ${\cal E}$ is, formally, the bundle corresponding to
the (trivial) deformation $B = \varepsilon I$. By inspection,
\begin{displaymath}
h(x) \: = \: x + \varepsilon \sigma_{(1)},
\end{displaymath}
and $h$ extends to higher symmetric powers of $S^* \otimes S$ in the
obvious way.

In the ordinary cohomology of $G(k,n)$, the only nonzero cohomology
classes correspond to Young diagrams inside a $k \times (n-k)$ box.
Young diagrams extending outside of that box should correspond to
vanishing cohomology classes, and define relations in the cohomology
ring.  To that end, we will consider relations defined by Young
diagrams with one row of boxes, extending outside of the $k \times
(n-k)$ box.

To that end, when $B = \varepsilon I$, we claim that
\begin{equation}\label{h(sigma)}
h(\sigma_{(r)}) = \sum_{i=0}^r {k+r-1 \choose i} \sigma_{(r-i)}
\sigma_{(1)}^i \varepsilon^i .
\end{equation}

For notational reasons, as we will be mixing Schur polynomials
$\sigma_{\mu}$ corresponding to Young diagrams and Coulomb branch
basis elements $\sigma_a$, which could become confusing, in this
appendix we will use the notation $x_a$ rather than $\sigma_a$ for
Coulomb basis elements to help distinguish the two.

Then,
\[
\sigma_{(r)} = \sum_{\alpha_1+\alpha_2+\cdots +\alpha_k = r}
x_1^{\alpha_1} x_2^{\alpha_2} \cdots x_k^{\alpha_k},
\]
\[
\begin{split}
h(\sigma_{(r)})&=\sum_{\alpha_1+\alpha_2+\cdots +\alpha_k = r}
(x_1+\varepsilon \sigma_{(1)})^{\alpha_1}
\cdots (x_k+\varepsilon \sigma_{(1)})^{\alpha_k},\\
&=\sum_{\alpha_1+\alpha_2+\cdots +\alpha_k = r} \left[
\left(\sum_{i_1=0}^{\alpha_1}{\alpha_1 \choose i_1}
x_1^{\alpha_1-i_1} \sigma_{(1)}^{i_1} \varepsilon^{i_1} \right)
\cdots \left(\sum_{i_k=0}^{\alpha_k}{\alpha_k \choose i_k}
x_k^{\alpha_k-i_k} \sigma_{(1)}^{i_k} \varepsilon^{i_k} \right)
\right],\\
&=\sum_{\alpha_1+\alpha_2+\cdots +\alpha_k = r}
\sum_{i_1=0}^{\alpha_1}\cdots \sum_{i_k=0}^{\alpha_k} {\alpha_1
\choose i_1} \cdots {\alpha_k \choose i_k} x_1^{\alpha_1-i_1} \cdots
x_k^{\alpha_k-i_k} \sigma_{(1)}^{i_1+\cdots+i_k}
\varepsilon^{i_1+\cdots+i_k} .
\end{split}
\]
The coefficient of $\varepsilon^i \sigma_{(1)}^i$, which we denote
by $g_i$, is
\[
\begin{split}
g_i =&\sum_{\alpha_1+\alpha_2+\cdots +\alpha_k = r}~
\sum_{i_1+\cdots+i_k = i} {\alpha_1 \choose i_1} \cdots {\alpha_k
\choose i_k}
x_1^{\alpha_1-i_1} \cdots x_k^{\alpha_k-i_k},\\
=&\sum_{\beta_1+\beta_2+\cdots +\beta_k = r-i}~ \sum_{i_1+\cdots+i_k
= i} {\beta_1+i_1 \choose i_1} \cdots {\beta_k+i_k \choose i_k}
x_1^{\beta_1} \cdots x_k^{\beta_k}.
\end{split}
\]
From the combinatorial formula
\begin{displaymath}
\sum_{i_1+\cdots+i_k = i} {\beta_1+i_1 \choose i_1} {\beta_2+i_2
\choose i_2} \cdots {\beta_k+i_k \choose i_k} =
{\beta_1+\beta_2+\cdots +\beta_k +k-1+i \choose i},
\end{displaymath}
we get
\begin{displaymath}
\begin{split}
g_i&=\sum_{\beta_1+\beta_2+\cdots +\beta_k = r-i}
{\beta_1+\beta_2+\cdots +\beta_k +k-1+i \choose i}
x_1^{\beta_1} \cdots x_k^{\beta_k},\\
&= {r-i+k-1+i \choose i} \sum_{\beta_1+\beta_2+\cdots +\beta_k =
r-i}
x_1^{\beta_1} \cdots x_k^{\beta_k},\\
&= {k+r-1 \choose i} \sigma_{(r-i)},
\end{split}
\end{displaymath}
and
\[
h(\sigma_{(r)})=\sum_{i=0}^r g_i \sigma_{(1)}^i \varepsilon^i
=\sum_{i=0}^r {k+r-1 \choose i} \sigma_{(r-i)} \sigma_{(1)}^i
\varepsilon^i .
\]

With some combinatorics one can then show
\begin{equation}\label{h(sigma_r)}
h(\sigma_{(r)}) \: = \: \sum_{j=0}^{k+r-n-1}
\sum_{i=j}^{\min\{n+j,r\}} \varepsilon^j {k+r-n-1 \choose j} I_{i-j}
\sigma_{(r-i)} \sigma_{(1)}^i .
\end{equation}

Now, we then claim that the kernel of the connecting map
\begin{displaymath}
\bigoplus_{r=0}^{k(n-k)}H^0_0\left(G(k,n),Sym^r(S^*\otimes S)\right)
\: \longrightarrow \: H^*(G(k,n),\wedge^* \mathcal{E}^\vee)
\end{displaymath}
can be generated by $R_{(r)}, r=n-k+1,n-k+2, \cdots$, where
\begin{displaymath}
R_{(r)} = \sum_{i=0}^{\min\{r,n\}} I_i \sigma_{(r-i)}
\sigma_{(1)}^i,
\end{displaymath}
with $I_i, i=0,1,\cdots,n$, the characteristic polynomials of $B$.

To show this, first note that the Giambelli formula
\begin{displaymath}
\sigma_\lambda = \det \left(
\begin{array}{cccc}
\sigma_{(\lambda_1)} & \sigma_{(\lambda_1+1)} & \cdots & \sigma_{(\lambda_1+k-1)} \\
\sigma_{(\lambda_2-1)} & \sigma_{(\lambda_2)} & \cdots & \sigma_{(\lambda_2+k-2)} \\
 & \multicolumn{2}{c}{\dotfill}\\
\sigma_{(\lambda_k+k-1)} & \sigma_{(\lambda_k+k)} & \cdots &
\sigma_{(\lambda_k)}
\end{array} \right)
\end{displaymath}
implies
\begin{equation}\label{Giambelli}
h(\sigma_\lambda) = h(\sigma_{(\lambda_1)}) F_1 +
h(\sigma_{(\lambda_1+1)}) F_2 + \cdots + h(\sigma_{(\lambda_1+k-1)})
F_k,
\end{equation}
where the $F_i$'s are polynomials of the $\sigma_{(j)}$'s. If
$h(\sigma_\lambda)$ is in the kernel of the connecting map, then
$\lambda_1 \geqslant  n-k+1$. Thus, $h(\sigma_{(r)}) ,
r=n-k+1,\cdots,k(n-k)$ generate the kernel. Furthermore, one can
write \eqref{h(sigma_r)} as
\begin{displaymath}
\begin{split}
h(\sigma_{(r)}) &= \sum_{j=0}^{k+r-n-1} \varepsilon^j {k+r-n-1
\choose j} \left(\sum_{i=0}^{\min\{n,r-j\}}
I_i \sigma_{(r-j-i)} \sigma_{(1)}^i\right) \sigma_{(1)}^j,\\
&= \sum_{j=0}^{k+r-n-1} \varepsilon^j {k+r-n-1 \choose j}
\sigma_{(1)}^j R_{r-j} ,
\end{split}
\end{displaymath}
for $r=n-k+1,\cdots,k(n-k)$, which implies that the $R_{(r)}$'s
generate the kernel, as claimed.

\end{document}